\documentclass[11pt, oneside]{scrartcl}   	
\usepackage{geometry}                		
\usepackage{latexsym}
\usepackage{graphicx}
\usepackage{amsmath}
\usepackage{amssymb}
\usepackage{slashed}
\usepackage{color}
\usepackage[dvipsnames, table]{xcolor}
\usepackage{mathtools}
\usepackage{siunitx}
\usepackage{xspace}
\usepackage[colorlinks=true,a4paper=true,backref=false,linktocpage=true,citecolor=blue,urlcolor=blue,linkcolor=blue,pdfpagemode=UseOutlines]{hyperref}
\newcommand{\gcl}{\cellcolor[gray]{0.90}}
\usepackage{multirow}

\newcommand{\be}{\begin{equation}}
\newcommand{\ee}{\end{equation}}
\newcommand{\bea}{\begin{eqnarray}}
\newcommand{\eea}{\end{eqnarray}}
\newcommand{\<}{\langle}
\renewcommand{\>}{\rangle}
\newcommand{\mc}{\mathcal}

\newcommand{\nn}{\nonumber}

\newcommand{\mmy}{\mu \mu \gamma}

\newcommand{\eps}{\epsilon}
\newcommand{\vareps}{\varepsilon}
\newcommand{\la}{\lambda}

\newcommand{\Bs}{\ensuremath{B^0_s}\xspace}
\newcommand{\Ds}{\ensuremath{D_s}\xspace}

\newcommand{\bBs}{\ensuremath{\bar B^0_s}\xspace}

\newcommand{\im}{{\rm Im}}

\newcommand{\DE}{\rm DE}
\newcommand{\Brems}{\rm Brems}
\newcommand{\GeV}{{\rm GeV}}
\newcommand{\MeV}{{\rm MeV}}

\newcommand{\Bsmmy}{\ensuremath{B_s \to \mu^+ \mu^- \gamma}\xspace}
\newcommand{\mm}{\ensuremath{\mu\mu}\xspace}
\newcommand{\Bsmm}{\ensuremath{B_s \to \mm}\xspace}
\newcommand{\boldBsmmy}{\ensuremath{\boldsymbol{B_s \to \mmy}}\xspace}

\newcommand{\Bstog}{\ensuremath{B_s \to \gamma}\xspace}

\newcommand{\FTA}{F_{TA}}
\newcommand{\FTV}{F_{TV}}

\newcommand{\FTVA}{\ensuremath{T_{\perp,\parallel}}}
\newcommand{\FVA}{\ensuremath{V_{\perp,\parallel}}\xspace}

\def\bsmumugamma{\ensuremath{\Bs \to \mu^+ \mu^- \gamma}\xspace}
\def\bsmumu{\ensuremath{\Bs \to \mu^+ \mu^-}\xspace}

\newcommand{\PP}{{\tt PP}\xspace}
\newcommand{\PE}{{\tt PE}\xspace}

\newcommand{\bra}[1]{\ensuremath{\langle #1|}}             
\newcommand{\ket}[1]{\ensuremath{|#1\rangle}}              

\DeclareOldFontCommand{\rm}{\normalfont\rmfamily}{\mathrm}
\DeclareOldFontCommand{\sf}{\normalfont\sffamily}{\mathsf}
\DeclareOldFontCommand{\tt}{\normalfont\ttfamily}{\mathtt}
\DeclareOldFontCommand{\bf}{\normalfont\bfseries}{\mathbf}
\DeclareOldFontCommand{\it}{\normalfont\itshape}{\mathit}
\DeclareOldFontCommand{\sl}{\normalfont\slshape}{\@nomath\sl}
\DeclareOldFontCommand{\sc}{\normalfont\scshape}{\@nomath\sc}

\definecolor{dartmouthgreen}{rgb}{0.05, 0.5, 0.06}

\newenvironment{Tabular}[2][1]
  {\def\arraystretch{#1}\tabular{#2}}
  {\endtabular}

\begin{document}

\begin{flushright}
\small
LAPTH-006/23
\end{flushright}
\vskip0.5cm

\begin{center}
{\sffamily \bfseries \LARGE \boldmath
\boldmath From $D_s \to \gamma$ in lattice QCD\\[0.2cm]to $B_s \to \mu \mu \gamma$ at high $q^2$
}\\[0.8 cm]
{\normalsize \sffamily \bfseries Diego Guadagnoli$^1$, Camille Normand$^1$, Silvano Simula$^2$, Ludovico Vittorio$^1$} \\[0.5 cm]
\small
$^1${\em LAPTh, Universit\'{e} Savoie Mont-Blanc et CNRS, Annecy, France}\\
[0.1cm]
$^2${\em INFN, Sezione di Roma Tre, Via della Vasca Navale 84, 00146 Rome, Italy}
\end{center}

\medskip

\begin{abstract}
\noindent We use a recent lattice determination of the vector and axial $D_s \to \gamma$ form factors at high squared momentum transfer $q^2$ to infer their $B_s \to \gamma$ counterparts. To this end, we introduce a phenomenological approach summarized as follows.
First, we describe the lattice data with different fit templates motivated by vector-meson dominance, that is expected to hold in the high-$q^2$ region considered. We identify reference fit ansaetze with one or two physical poles, that we validate against alternative templates.
Then, the pole residues can be unambiguously related to the appropriate couplings involving the pseudoscalar, the vector mesons concerned, and the photon---or tri-couplings---and the latter can be expressed as sums over quark magnetic moments, weighed by their e.m. charges.
This description obeys a well-defined heavy-quark scaling, that allows to parametrically scale up the form factors to the $B_s \to \gamma$ case.
We discuss a number of cross-checks of the whole approach, whose validation rests ultimately in a first-principle determination, e.g. in lattice QCD. Finally, we use our obtained form factors to reassess the SM prediction of $\mc B(B_s \to \mu^+ \mu^- \gamma)$ in the range $\sqrt{q^2} \in [4.2, 5.0]$ GeV, where an experimental measurement is awaited.
\end{abstract}

\date{}							

%


\section{Introduction} \label{sec:intro}

Recently, the LHCb collaboration set a first limit on the rare-and-radiative decay \bsmumugamma for $q^2 > (4.9~\GeV)^2$~\cite{LHCb:2021vsc,LHCb:2021awg} via the ``indirect'' method of extracting this decay as a shoulder of $\bsmumu$ \cite{Dettori:2016zff}. This comes with several advantages: it allows to use the established di-muon trigger rather than a dedicated one; it avoids an inefficient photon detection and reconstruction; it measures the decay in the high-$q^2$ region, which is largely immune from resonance pollution, is mostly sensitive to semi-leptonic Wilson coefficients, as opposed to dipole operators, and is the best accessible to lattice-QCD simulations aimed at first-principle determinations of the necessary $B_s \to \gamma$ form factors (FFs).

The amplitude for this decay arises from two distinct sets of contributions. In a first set the required e.m. current $J_{\rm e.m.}$ acts on the final-state dimuon, and it does not enter the matrix element involving the initial $B_s$, the final $\gamma$ and the weak-transition operator $\mc O_w$. This contribution is often denoted as ``final-state radiation'' (FSR). The second set of contributions involves the $T$-product of $J_{\rm e.m.} \mc O_w$, and may be referred to as ``initial-state radiation'' (ISR).\footnote{This component includes photon emission by the Standard-Model d.o.f. integrated out at the weak scale. Hence the ISR designation is not very accurate. It is intended to designate anything other than FSR.} The main points are \cite{Dettori:2016zff} that interference between the ISR and FSR amplitudes is completely negligible in the full kinematic range; that ISR and FSR dominate the spectrum in two separate regions of $q^2$, respectively below and above $(5~\GeV)^2$; and that FSR is well-understood, and {\em subtracted} from the measured \bsmumu rate. As a consequence, ISR gives rise to a well-defined observable. In particular, LHCb's current limit~\cite{LHCb:2021vsc,LHCb:2021awg}
\be
\label{eq:Bsmmy_exp}
\mc B(\bsmumugamma)[q^2 > (4.9~\GeV)^2] < 2.0 \times 10^{-9}~,
\ee
effectively covers only the limited range $\sqrt{q^2} \in [4.9, 5.0]~\GeV$, whilst the measurement could go as low as $4.2~\GeV$---below which broad-charmonium pollution becomes non-negligible. Denoting the observable in this range as $\mc B(\bsmumugamma)[4.2, 5.0]$, we remark that its SM prediction ranges from $2\times 10^{-10} \div 3 \times 10^{-10}$ with the quark-model FFs of Refs. \cite{Melikhov:2004mk, Kozachuk:2017mdk} to a figure about one order of magnitude (o.o.m.) larger with the LCSR FFs of Ref. \cite{Janowski:2021yvz}. This larger prediction would suggest that the next $\bsmumugamma$ update by LHCb with the same ``indirect'' method may well be a {\em measurement}, not a limit. Further theoretical work to narrow down such range of SM prediction is thus urgently required to back a possibly imminent update of eq. (\ref{eq:Bsmmy_exp}).

Due to the fortunate circumstance of negligible resonance pollution in our range of interest $\sqrt{ q^2} \in [4.2, 5.0]~\GeV$, the SM prediction rests entirely on controlling the $B_s \to \gamma$ FFs. The dominant ones in this region are $F_V$ and $F_A$, defined as \cite{Kruger:1996cv}
\bea
\label{eq:FV_FA_defs}
\< \gamma(k,\vareps) |\bar s \gamma_\mu \gamma^5 b| \bar B(p) \> &=& i e \left( \vareps^*_\mu  p\cdot k - k_\mu \vareps^* \cdot p\right) \frac{F_A(q^2)}{m_{\Bs}}~,\nn \\
\< \gamma(k,\vareps) |\bar s \gamma_\mu b| \bar B(p) \> &=& e \, \eps_{\mu \vareps^* p k} \,\frac{F_V(q^2)}{m_{\Bs}}~,
\eea
where $p,k$ denote momenta, with $q^2 = (p-k)^2$, $\vareps$ a polarization vector, and $\eps$ the antisymmetric tensor.\footnote{Eq. (\ref{eq:FV_FA_defs}) assumes the convention $\eps_{0123} = +1$.}

The $F_{V,A}$ FFs have been calculated in a handful of works. They include the mentioned Ref.~\cite{Melikhov:2004mk}, that uses a relativistic quark model, updated in Ref.~\cite{Kozachuk:2017mdk} and to be referred to as KMN; Ref.~\cite{Beneke:2020fot} (BBW), using soft-collinear effective theory and rigorous factorization methods, thereby valid for $q^2 < 6~\GeV^2$ only, and of course outside the region dominated by the $\phi$ resonance; Ref. \cite{Janowski:2021yvz} (JPZ), which uses a light-cone sum-rules (LCSR) approach. Note that KMN and JPZ results are stated to be reliable for every $q^2$ in the physical region, although in either case the main constraints are defined for low $q^2$. Comparing JPZ with KMN FFs, one finds sizeable, several-100\%, differences,\footnote{A low-$q^2$ comparison (including the respective errors) between JPZ and BBW may be found in Ref. \cite{Carvunis:2021jga}.} leading to the mentioned o.o.m. difference between the corresponding branching-ratio predictions.  

The goal of this paper is to reappraise the high-$q^2$ FFs and the ensuing branching-ratio prediction. To this end, we start from the recent Lattice QCD (LQCD) computation \cite{Desiderio:2020oej} of the hadronic FFs entering in radiative $P \to \gamma \ell \nu$ decays, where $P= \pi,\,K,\,D,\,D_s$.\footnote{A new study of lattice-QCD methods to determine the FFs for radiative leptonic decays of pseudoscalar mesons has appeared very recently \cite{Giusti:2023pot}.} For our purpose, we will be mainly interested in the $D_s$-meson case. Concerning $D_s \to \gamma \ell \nu$ decays, Ref.~\cite{Desiderio:2020oej} provides direct access to the region of low $x_{\gamma}$ in the range $0.05 \lesssim x_{\gamma} \lesssim 0.4$, where
\be
\label{eq:xgamma}
x_{\gamma} \equiv \frac{2 p \cdot k}{m_{D_s}^2} = 1- \frac{q^2}{m_{D_s}^2}~.
\ee
Low $x_{\gamma}$ corresponds to high $q^2$. In particular, note that the specific high-$q^2$ region $[4.2, 5.0]^2$ $\GeV^2$ of interest for the $B_s \to \mu \mu \gamma$ measurement of Ref. \cite{Dettori:2016zff} corresponds to $x_\gamma \in [0.39, 0.13]$, which neatly overlaps with the range covered by Ref.~\cite{Desiderio:2020oej} for $D_s \to \gamma \ell \nu$ decays. Our purpose is therefore to estimate the $B_s \to \gamma$ counterpart of Ref.~\cite{Desiderio:2020oej}, awaiting a direct determination of the same on the lattice. Our estimation will be based on heavy-quark scaling arguments and will be used for a new determination of our observable of interest, namely $\mc B(\bsmumugamma)[4.2, 5.0]$, within the SM.

The paper is organized as follows. In Section \ref{sec:Ds_FFs} we analyze the LQCD data for $D_s \to \gamma \ell \nu$ decays in Ref.~\cite{Desiderio:2020oej}. Here we discuss different functional ansaetze to fit the data in our range of interest, and establish a connection with the continuum and chiral extrapolations described at the end of Section V of Ref.~\cite{Desiderio:2020oej}. In Section \ref{sec:FF_extrapolation} we present our phenomenological approach to extrapolate our $D_s$-sector results to the $B_s$ case. Here we also compare our findings with other FF results in literature. In Section \ref{sec:Bsmumuy_prediction} we discuss our SM prediction of $\mc B(\bsmumugamma)[4.2, 5.0]$. Finally, Section \ref{sec:conclusions} presents our conclusions.

\section{\boldmath The hadronic form factors in $D_s \to \gamma \ell \nu$ decays}
\label{sec:Ds_FFs}

\subsection{Introduction}

Our stated aim is to compute the hadronic $B_s \to \gamma$ FFs in the high-$q^2$ region, which corresponds to a low photon energy. This supports a parametrization of the FFs inspired by Vector Meson Dominance (VMD). Following Ref.\,\cite{Becirevic:2009aq} and focusing, for instance, on the vector case, we have
\be
\label{eq:BHKeq}
\< \gamma |\bar s \gamma_\mu b| \bar B_s \> \simeq \sum_{\lambda} \frac{\< 0 |\bar s \gamma_\mu b| B_s^* (\vareps_{\lambda}) \>\<  B_s^* (\vareps_{\lambda}) | B_s \gamma \>}{q^2 - m_{B_s^*}^2}~,
\ee
where the r.h.s. matrix elements are defined as
\bea
\< 0 |\bar s \gamma_\mu b| B_s^* (\vareps_{\lambda}) \> &=& \vareps_{\mu}^{\lambda} m_{B_s^*} f_{B_s^*}~,\nn \\
\< B_s (p^{\prime}) \gamma(p,\,\vareps_{\lambda^{\prime}}) | B_s^* (q,\,\vareps_{\lambda})  \> &=& e \, \epsilon_{\eta \vareps q p^{\prime}} g_{B_s^* B_s \gamma}~.
\eea
with analogous formul\ae\ for the axial-vector case~\cite{Becirevic:2009aq}.

Eq.\,(\ref{eq:BHKeq}) plus the axial-vector-channel analogue may then be applied to the dispersion representation of the hadronic FFs. For comparison with other results present in the literature, let us mention the alternative notation \cite{Janowski:2021yvz}
\be
\label{eq:diffnotFFs}
V_{\perp}(q^2) = - F_V(q^2)~, \qquad V_{\parallel}(q^2) = - F_A(q^2)~,
\ee
where the FFs $F_{V,A}(q^2)$ were introduced in eq.\,(\ref{eq:FV_FA_defs}). Focusing again on the vector case for simplicity, we have
\be
\label{eq:FF=disprelBV}
V_{\perp} (q^2) = \frac{1}{\pi} \int_{0}^{\infty} dt \frac{\im[V_{\perp} (t)]}{t-q^2} = \frac{r_{\perp}}{1-q^2/m_{B_s^*}^2} + ...
\ee
where the dots represent one- as well as multi-particle contributions from states heavier than the $B_s^*$. Making use of eq.\,(\ref{eq:BHKeq}), we can then relate the residue $r_{\perp}$to the ``tri-coupling'' $g_{B_s^* B_s \gamma}$ as
\be
\label{eq:res=gBV}
r_{\perp} = \frac{m_{B_s} f_{B_s^*}}{m_{B_s^*}} g_{B_s^* B_s \gamma}~.
\ee
A parameterization of the $V_{\parallel} (q^2)$ FF analogous to eq. (\ref{eq:FF=disprelBV}), with resonant mass $m_{B_{s1}}$ and residue $r_{\parallel}$, leads to  
\be
\label{eq:res=gBA}
r_{\parallel} = \frac{m_{B_s} f_{B_{s1}}}{m_{B_{s1}}} g_{B_{s1} B_s \gamma}~.
\ee

\subsection{\boldmath LQCD data for $D_s \to \gamma \ell \nu$ decays and VMD ansatz}

\subsubsection{Basic application to \boldmath $V_{\perp}^{D_s}$}

Let us consider the LQCD data in Ref.\,\cite{Desiderio:2020oej} for $D_s \to \gamma$ decays, which have been directly computed in the region $0.05 \lesssim x_{\gamma} \lesssim 0.4$, $i.e.$ at low $x_{\gamma}$. We analyze these data by extending to the $D_s$ sector the ansaetze in eqs.\,(\ref{eq:FF=disprelBV})-(\ref{eq:res=gBA}). In what follows, we will then refer to the hadronic FFs in the $D_s$ sector as $V_{\perp,\parallel}^{D_s}(q^2)$, which we parameterize as
\be
\label{eq:FF=disprel}
V_{\perp[\parallel]}^{D_s} (q^2) = \frac{1}{\pi} \int_{0}^{\infty} dt \frac{\im[V_{\perp[\parallel]}^{D_s} (t)]}{t-q^2} = \frac{r_{\perp[\parallel]}^{D_s^* [D_{s1}]}}{1-q^2/m_{D_s^*\,[D_{s1}]}^2} + ...
\ee
where the residues $r$ are related to the tri-couplings via the relations
\be
\label{eq:res=g}
r_{\perp}^{D_s^*} = \frac{m_{D_s} f_{D_s^*}}{m_{D_s^*}} g_{D_s^* D_s \gamma},\,\qquad r_{\parallel}^{D_{s1}} = \frac{m_{D_s} f_{D_{s1}}}{m_{D_{s1}}}g_{D_{s1} D_s \gamma}.
\ee
We then use the LQCD data in Ref.\,\cite{Desiderio:2020oej} to infer the numerical values of the residues, which we then translate into predictions for the tri-couplings $g_{D_s^* D_s \gamma}$ and $g_{D_{s1} D_s \gamma}$. We deem this exercise  instructive also in view of a direct comparison of our results with the literature available. We are aware of two other estimates\footnote{In principle one could envisage a direct determination from experiment \cite{ParticleDataGroup:2022pth}, but the $D_s^*$ lifetime is unfortunately not available.} of $g_{D_s^* D_s \gamma}$: the direct determination by HPQCD \cite{Donald:2013sra}
\be
\label{eq:triHPQCD}
g_{D_s^* D_s \gamma} = 0.10 (2)~{\rm GeV}^{-1}~,
\ee
as well as the LCSR computation \cite{Pullin:2021ebn}
\be
\label{eq:triLCSR}
g_{D_s^* D_s \gamma} = 0.60^{+0.19}_{-0.18}~{\rm GeV}^{-1}~.
\ee

\begin{figure}[h!]
 \centering
 \includegraphics[width=0.49\textwidth]{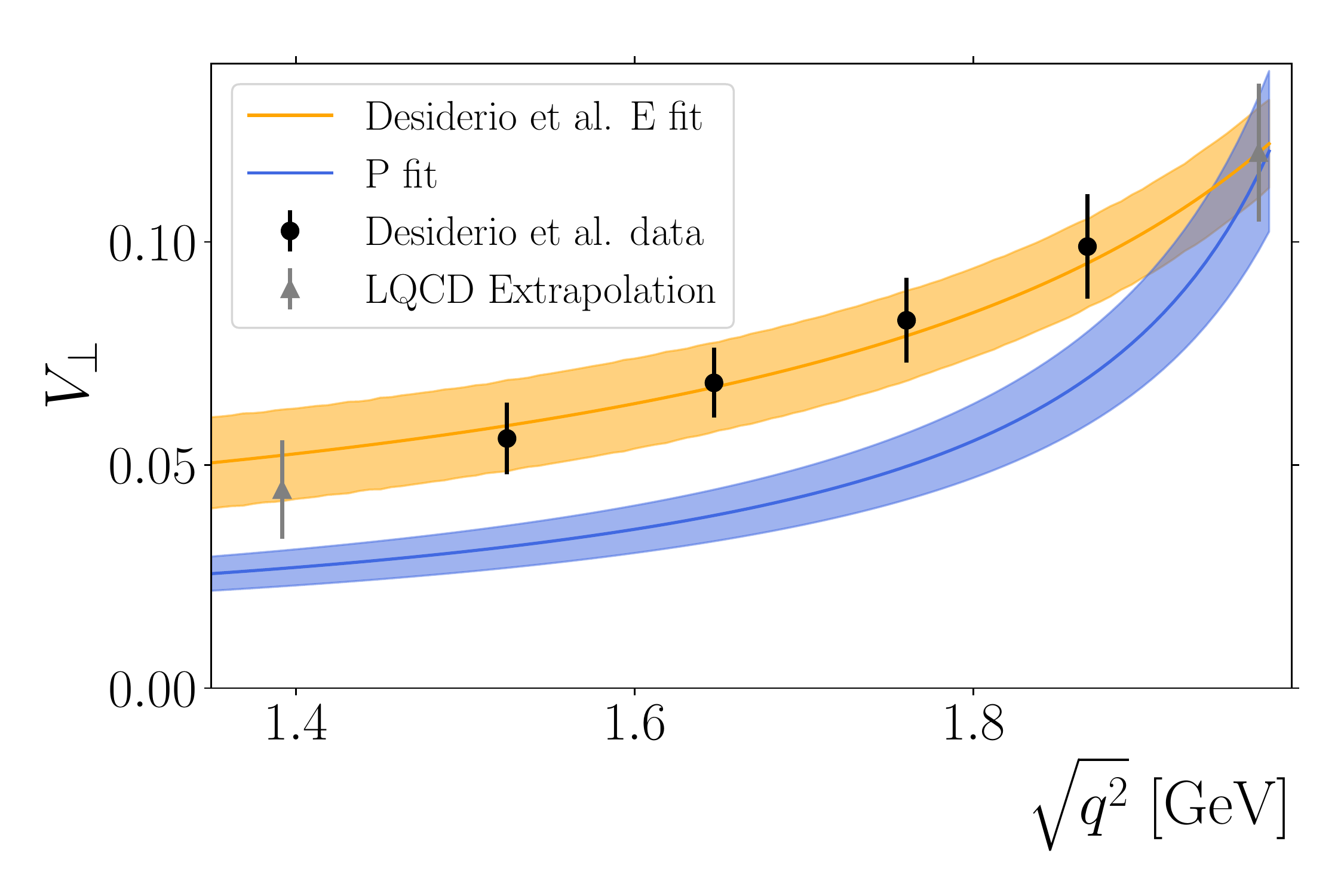}
 \includegraphics[width=0.49\textwidth]{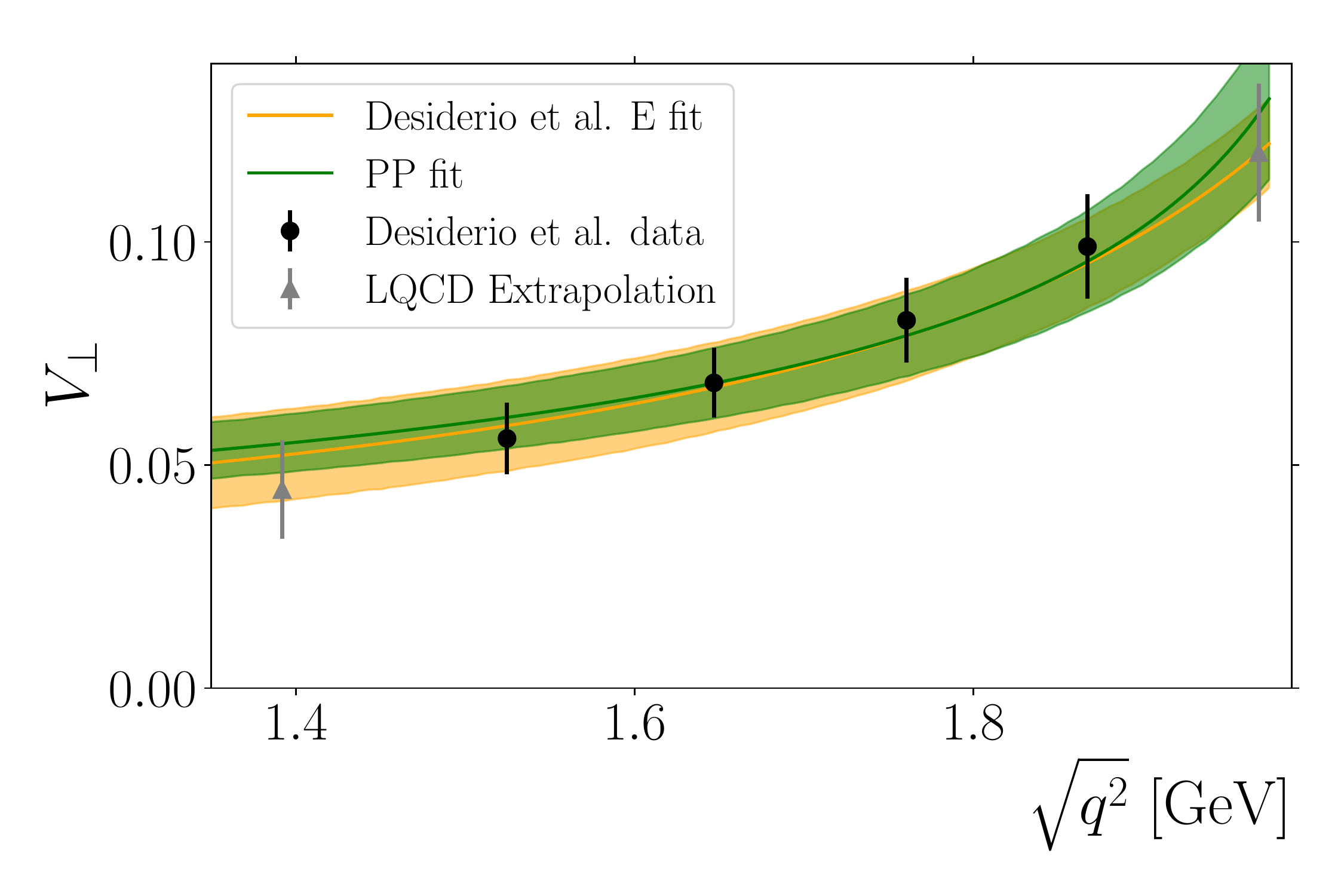}
\\
 \includegraphics[width=0.49\textwidth]{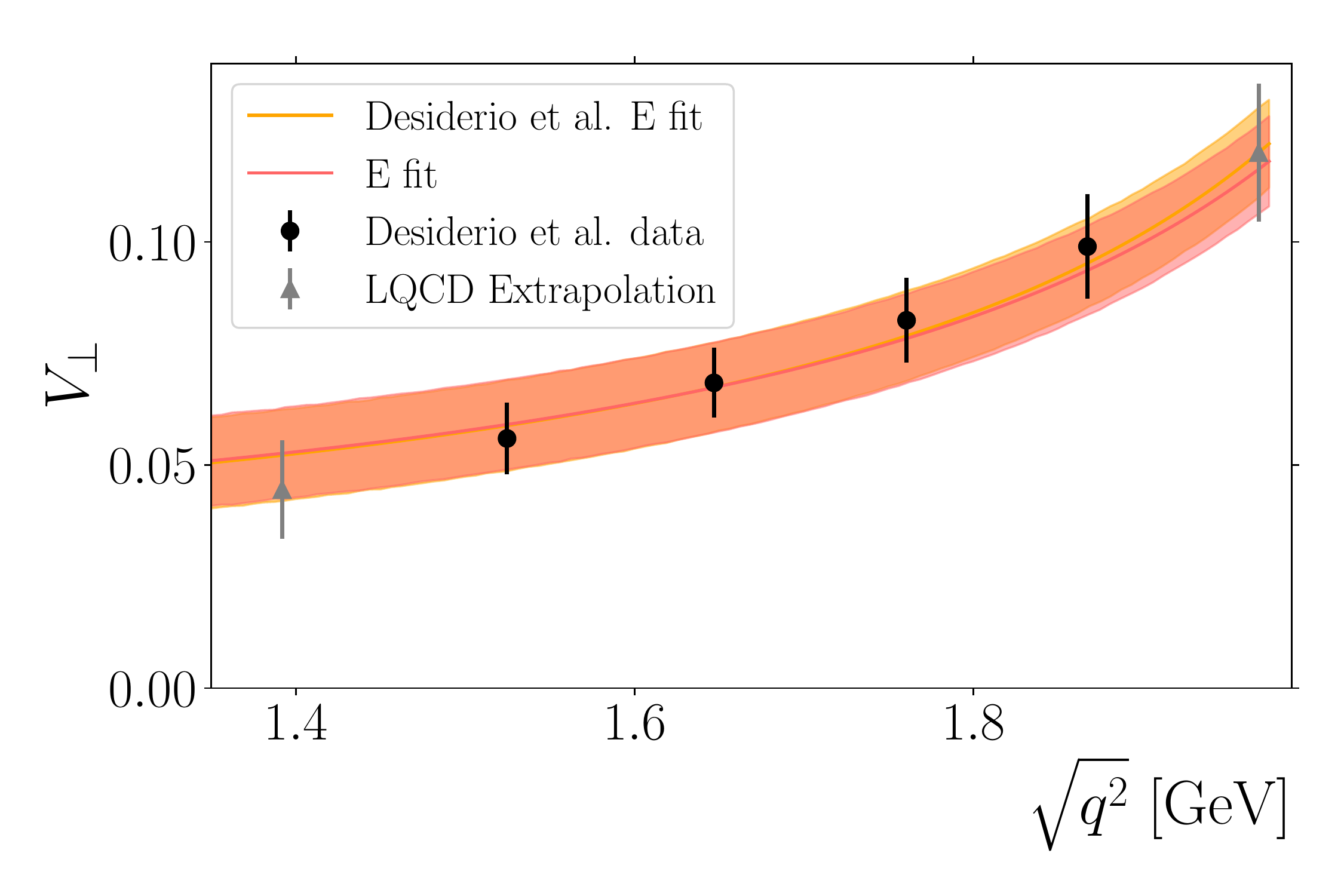}
 \includegraphics[width=0.49\textwidth]{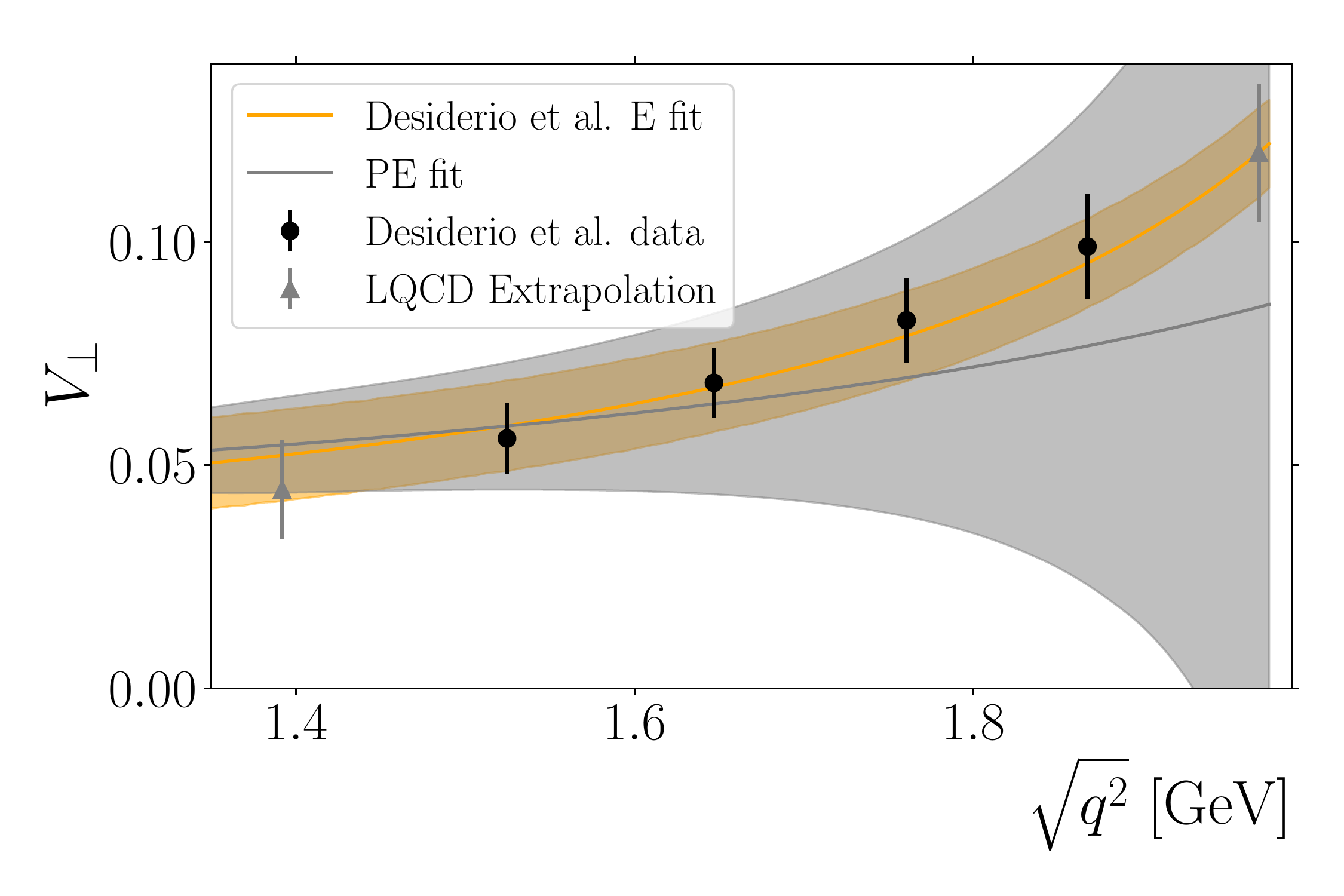}
 \centering
\caption{Form factor $V_{\perp}^{D_s}   (\sqrt{q^2})$ vs. $\sqrt{q^2}$. The black vs. grey points are the data from Ref.~\cite{Desiderio:2020oej}, respectively directly computed in LQCD, or extrapolated. Only the directly computed data are used as inputs of our fits, while the extrapolated data are shown for completeness (see text for further details). The orange band is the result of the pole-like fit in the end of Section V of Ref.~\cite{Desiderio:2020oej}. From top left to bottom right: {\tt P} fit (blue), eq.\,(\ref{eq:FV_Pfit}); {\tt PP} fit (green), eq.\,(\ref{eq:FV_PPfit}); {\tt E} fit (red), eq.\,(\ref{eq:FV_Efit}); {\tt PE} fit (grey), eq.\,(\ref{eq:FV_PEfit}).}
\label{fig:FV_fits}
\end{figure}

The ansatz in eq.\,(\ref{eq:FF=disprel})---with only the first term on the r.h.s.---yields the following value for the residue $r_{\perp}^{D_s^*}$
\be
\label{eq:FV_Pfit}
r_{\perp}^{D_s^*} = 0.015 (2),\,\qquad \chi^2 = 32, \,\qquad \mbox{$p$-value}<10^{-6}~.
\ee
Note that this value is lower than the residues that could be inferred from the tri-couplings in eqs.\,(\ref{eq:triHPQCD})-(\ref{eq:triLCSR}), namely $r_{\perp,\,\rm{HPQCD}}^{D_s^*} = 0.025\,(5)$ and $r_{\perp,\,\rm{LCSR}}^{D_s^*} = 0.15\,(5)$, respectively. The result in eq.~(\ref{eq:FV_Pfit}) uses only the data in the region $x_{\gamma} \in [0.1,0.4]$ (including their correlations) as inputs of our fit. Unless specified otherwise, this range will be our reference one throughout our analysis, because the corresponding data are directly computed on (rather than extrapolated from) the lattice, and because this may be the most conservative range for VMD to hold---i.e. we expect VMD to hold to a lesser degree in a larger range. The numerical value of the residue in eq.\,(\ref{eq:FV_Pfit}) generates the blue band of fig.\,\ref{fig:FV_fits} (top left), where the LQCD data are also shown for comparison. As is visually evident, and as also shown by the $p$-value in eq.\,(\ref{eq:FV_Pfit}), the blue band does not reproduce at all the LQCD data used as inputs.

\subsubsection{Discussion of alternative fitting ansaetze}

The fitting template described in eq.\,(\ref{eq:FF=disprel}) may be justified through the following underlying physical picture: close to the $q^2$ endpoint of the $D_s \to \gamma$ FFs, the $D_s \gamma$ system has a total invariant mass close to that of the first vector (the $D_{s}^*$) or axial (the $D_{s1}$) excited state---and the same quantum numbers. So the FF may plausibly follow a VMD ansatz \cite{Becirevic:2009aq}, as previously discussed. Concretely, the assumption in eq.\,(\ref{eq:res=g}) of a single, physical pole may be unrealistic, as also suggested by fig.~\ref{fig:FV_fits} (left panel).
Both in the $V_{\perp}^{D_s}$ and $V_{\parallel}^{D_s}$ channels, one may in particular consider the following ansaetze:
\begin{itemize}
\item[\tt P] A single, physical pole, corresponding to the first physical excited state. In this case, one fits for the residue alone, and the latter is related unambiguously to the tri-coupling~\cite{Becirevic:2009aq}. This {\tt P} ansatz is thus parameterized as
\be
\label{eq:P_ansatz}
V_{\chi}(q^2) = \frac{r_{\chi 1}}{1 - q^2/m^2_{{\rm ph}1}}~,~~~~\mbox{with }\chi = \perp,\parallel~,
\ee
and we fit for $r_{\chi 1}$ only, whilst $m_{\rm ph1}$ is a physical mass with known value.

\item[\tt PP] Like fit {\tt P}, but including also a second, physical pole, to the extent that this second-excited-state mass is also known. This ansatz makes sense if the two excited states have both an invariant mass close to that of the $D_s \gamma$ system in the $q^2$ region covered by the lattice data we are using. This \PP ansatz is thus parameterized similarly as eq. (\ref{eq:P_ansatz}), but for two fitted residues $r_{\chi 1}$ and $r_{\chi 2}$, and two fixed masses $m_{{\rm ph}1}$ and $m_{{\rm ph}2}$. Both residues belong to physical poles, and thus can be related to tri-couplings via eq.~(\ref{eq:res=g}).

\item[\tt E] A single {\em effective} pole, with a fitted residue $r_{\chi 1}$ and a fitted mass $m_{\chi 1}$. This ansatz may provide an economic way to take into account the first resonance as well as the structure above it in a single pole parametrization. However, because an {\tt E} fit is not a physical pole (with namely a fixed physical mass), the residue cannot be related to the tri-coupling of a particular meson state.

\item[\tt PE] Like fit {\tt P}, plus an {\em effective} pole, which would account for higher resonances or for the continuum. One would then fit for two additional parameters with respect to a {\tt P} fit, namely the residue and the pole of the effective pole term. This ansatz may be justified by a similar argument as for the {\tt PP} one---in particular, the continuum threshold may be estimated as $s_{\rm th} = (m_{D_s} + m_\rho)^2$. This \PE ansatz is thus parameterized as
\be
\label{eq:PE_ansatz}
V_{\chi}(q^2) = \frac{r_{\chi 1}}{1 - q^2/m^2_{{\rm ph}1}} + \frac{r_{\chi 2}}{1 - q^2/m^2_{\chi 2}}~,
\ee
where we fit for $r_{\chi 1}$, $r_{\chi 2}$, and the mass of the second pole $m_{\chi 2}$, whereas $m_{{\rm ph}1}$ is, again, fixed.

\item[\tt PPE] Like fit {\tt PP}, but including also a further, {\em effective} pole. In the conventions of the previous examples, such {\tt PPE} ansatz fits for the parameters $r_{\chi 1}$, $r_{\chi 2}$, $r_{\chi 3}$ and $m_{\chi 3}$.

\end{itemize}
\subsubsection{Application to \boldmath $V_{\perp}^{D_s}$}\label{sec:appl_FV}

In the nomenclature introduced in the previous section, \PP or \PE fits to the data may provide a more realistic description of $V_{\perp}^{D_s}(q^2)$. In fact, above the $D_s^*(2112)$ meson, there exists at least one further $1^-$ resonance, the $D_{s1}^*(2700)$ \cite{ParticleDataGroup:2022pth}, whose mass is just underneath the expected continuum threshold. The above two states may be well described through a \PP fit, which yields
\bea
\label{eq:FV_PPfit}
&r_{\perp 1} = 0.009 \pm 0.003~,~~~~ r_{\perp 2} = 0.029 \pm 0.005~,~~~~\rho(r_{\perp 1}~, r_{\perp 2}) = -0.44~,& \nn \\
&\chi^2=1.5~,~~~~ \mbox{$p$-value}=0.48 ~.&
\eea
We verify the consistency of these values with a \PE ansatz, where we replace the measured $m_{D_{s1}^*}$ with the parameter $m_{\perp 2}$, determined by the fit. We obtain
\bea
\label{eq:FV_PEfit}
&r_{\perp 1} =-0.00 (2)~,~~~~ r_{\perp 2} = 0.04 (2)~,~~~~ m_{\perp 2}= 2.7 (4)~\GeV~,&
\nn \\
&\rho(r_{\perp 1}~, r_{\perp 2}) = -0.97~,~~~~\rho(r_{\perp 1}~, m_{\perp 2}) = -0.63~,~~~~\rho(r_{\perp 2}, m_{\perp 2}) = +0.52~,&\nn \\
&\chi^2 = 2.0~.&
\eea
Throughout the text $\rho$ represents the correlation between the parameters in argument.
We note that either of the \PP and \PE fits give more weight to the second than to the first resonance, as is apparent from the residue values. In fact, one may further perform a \PP fit where the first residue is constrained by the HPQCD result in eq.\,(\ref{eq:triHPQCD}), via a Gaussian prior. This, however, results in a sizeable increase in the $\chi^2$ value. A further verification of the relative role of the two resonances may be obtained through an {\tt E} fit, akin to the phenomenological fit performed in Ref.\,\cite{Desiderio:2020oej}, but for the range of LQCD data considered. Our {\tt E} fit yields
\bea
\label{eq:FV_Efit}
&r_{\perp 1} = 0.034 (4)~,~~~~m_{\perp 1}=2.34 (7)\,\GeV~,~~~~\rho({r_{\perp 1}, m_{\perp 1}}) = -0.34~,&\nn \\
&\chi^2 = 1.0~.&
\eea
The outcome residue is close to the second residue of the \PP fit, and the outcome mass is about 10\% larger than the $D_s^*$'s. All of \PP, \PE and {\tt E} fits are shown in fig.~\ref{fig:FV_fits}. While the \PP- and {\tt E}-fit results are very similar to the pole-like fit in Ref.\,\cite{Desiderio:2020oej}, the \PE fit yields a larger uncertainty in the high $q^2$ region.\footnote{A natural question is whether the limited number of LQCD data used---as mentioned, only those directly calculated on the lattice, but including their correlations---may imply that the \PE fit is unreliable, the total number of degrees of freedom in the fit being one in this case. We have checked this possibility by repeating the \PE fit in the enlarged region $x_{\gamma} \in [0,0.6]$, with basically identical results than in eq.\,(\ref{eq:FV_PEfit}).}

In short, a {\tt P} fit does not describe the data well---see eq. (\ref{eq:FV_Pfit}) and blue band in fig.\,\ref{fig:FV_fits}, first panel. \PP or \PE fits provide a much improved description, but they give more weight to the second than to the first resonance (see residue values), this conclusion being confirmed by an {\tt E} ansatz. Remarkably, a physical interpretation of this result may be obtained within the model in Ref. \cite{Godfrey:2016nwn}, which suggests that the second pole of the vector channel has a larger decay width (to the ``ground-state'' pseudo-scalar meson plus the photon) than the first pole, as a result of a larger coupling and larger phase space, i.e. because of accidental reasons.\footnote{Note that this interpretation holds for the $B_{d,s}$ case. With the information in Ref.~\cite{Godfrey:2015dva} we are unable to pursue a similar interpretation for the $D_s$ case.} Another, more circumstantial explanation is the paucity of data we have at our disposal. In any case, the \PP, \PE or simply {\tt E} ansaetze provide mutually consistent results. Concretely, the \PP fit is equivalent to the pole-like fit performed in Ref.\,\cite{Desiderio:2020oej}, but we think it has the advantage of providing a plausible physical interpretation. We thus pick the \PP result as reference.

From eqs. (\ref{eq:FV_PPfit}) and (\ref{eq:res=gBV}), and using the numerical values of the meson decay constants in Appendix \ref{app:meson_decay_constants}, we can infer the following prediction for the tri-coupling associated with the first pole in the vector channel
\be
\label{eq:g_from_PP_in_FV}
g_{D_s^* D_s \gamma} = 0.04 (1)~{\rm GeV}^{-1}~.
\ee
The determinations in eqs.\,(\ref{eq:triHPQCD}) and (\ref{eq:g_from_PP_in_FV}) are in tension at the $\sim 2.5\sigma$ level. This may be due to the correlations among the LQCD data, which contain the information about the extrapolation of $V_{\perp}^{D_s}(x_{\gamma})$ at $x_{\gamma}=0$ performed in Ref.\,\cite{Desiderio:2020oej}. Note that correlations modify drastically the result on the residue---in particular an uncorrelated fit returns the same residue one would obtain with the HPQCD estimate of the tri-coupling in eq. (\ref{eq:triHPQCD}) and Ref.~\cite{Donald:2013sra}. Finally, both these values are sizeably lower than, and appear inconsistent with, the LCSR determination in eq.\,(\ref{eq:triLCSR}).

\subsubsection{Application to \boldmath $V_{\parallel}^{D_s}$}\label{sec:appl_FA}

\begin{figure}[h!]
 \centering
 \includegraphics[width=0.49\textwidth]{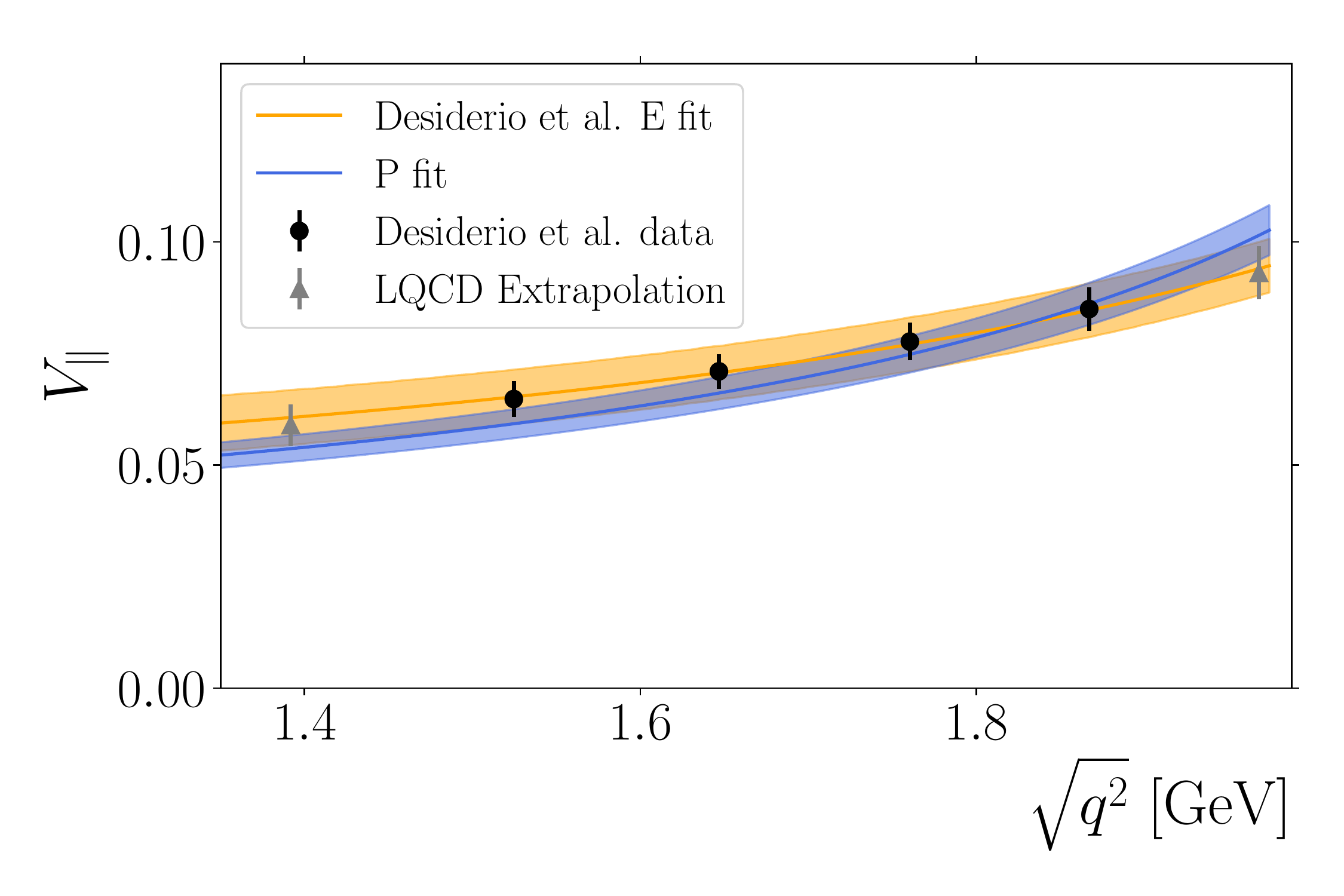}
 \includegraphics[width=0.49\textwidth]{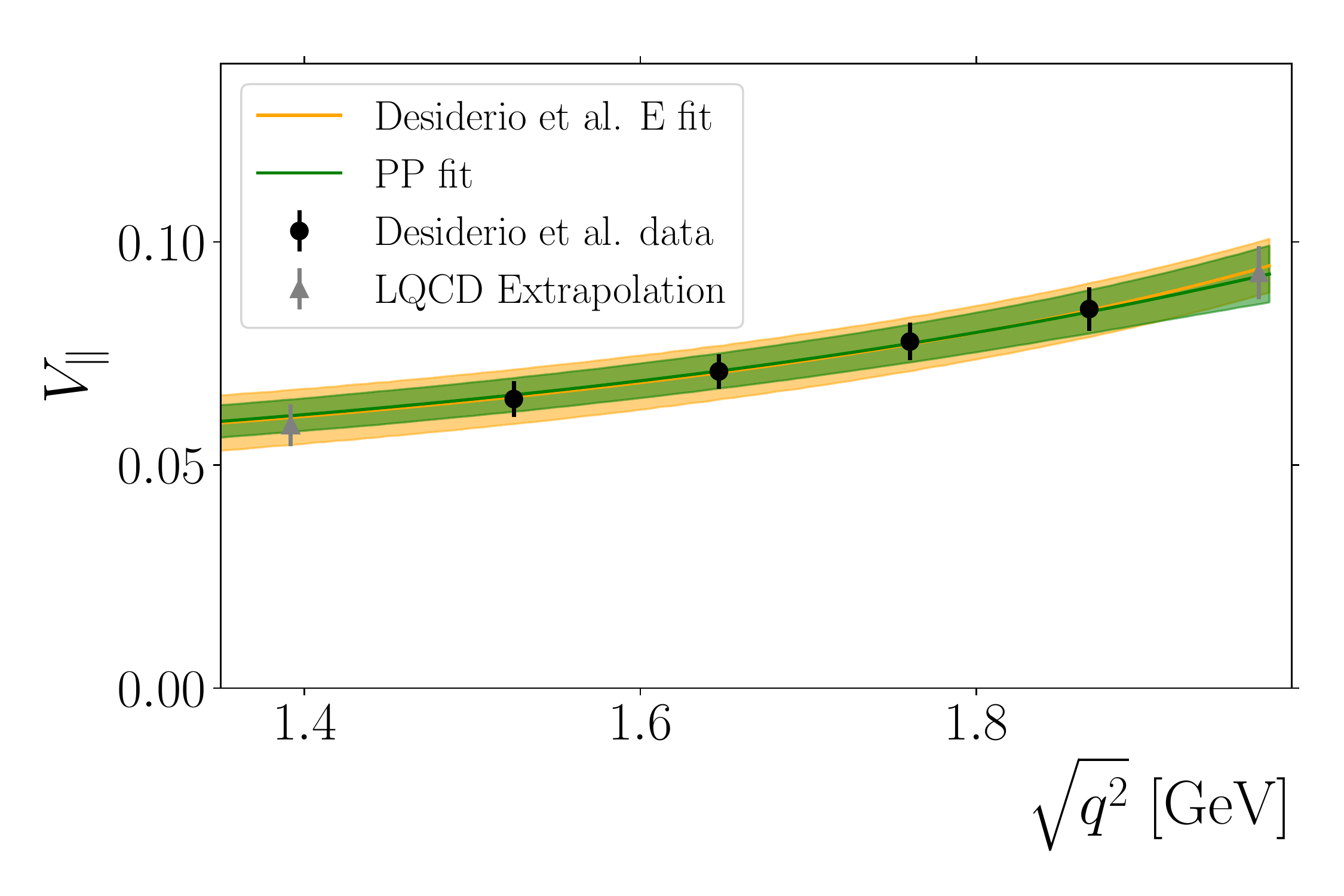}
\\
 \includegraphics[width=0.49\textwidth]{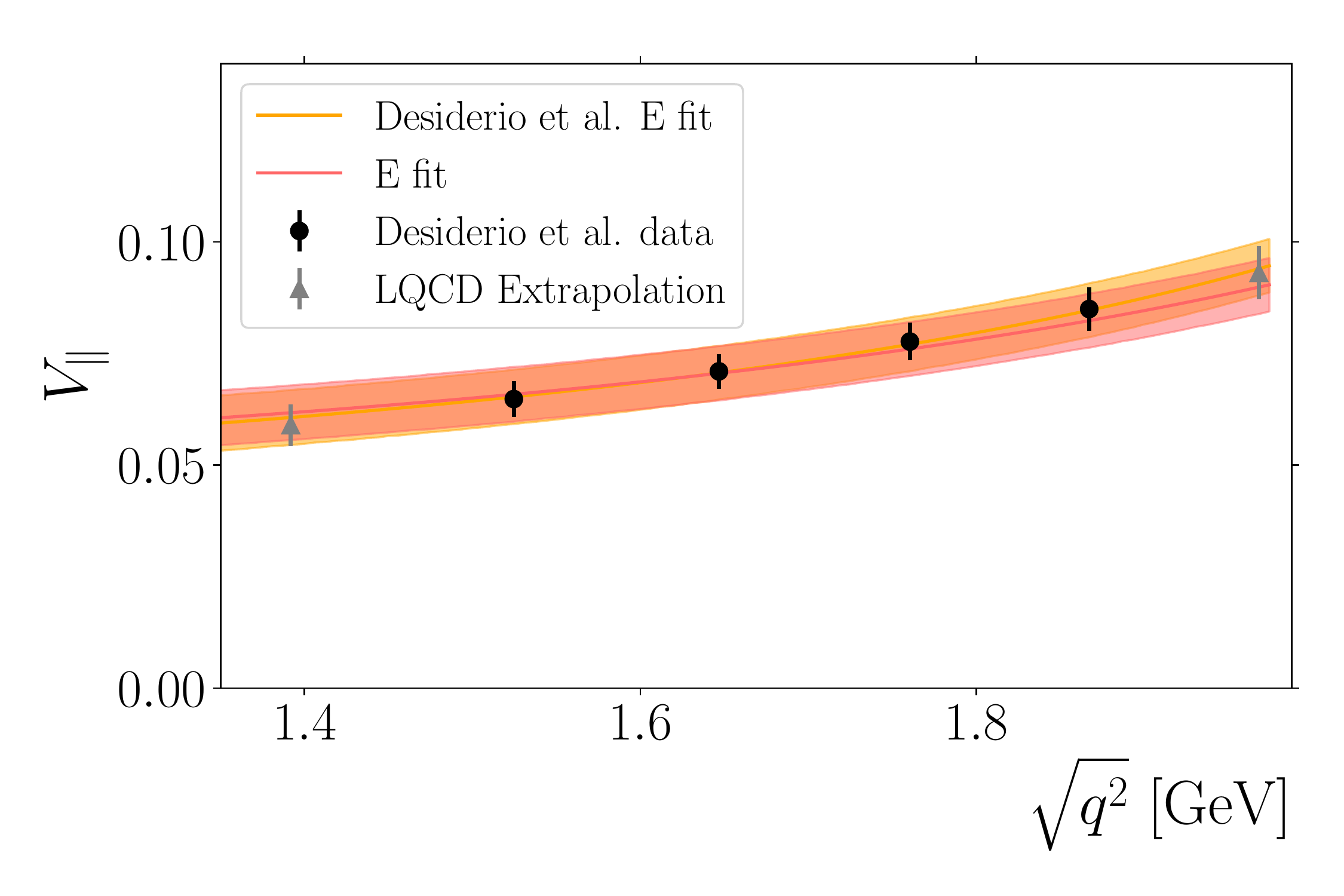}
 \includegraphics[width=0.49\textwidth]{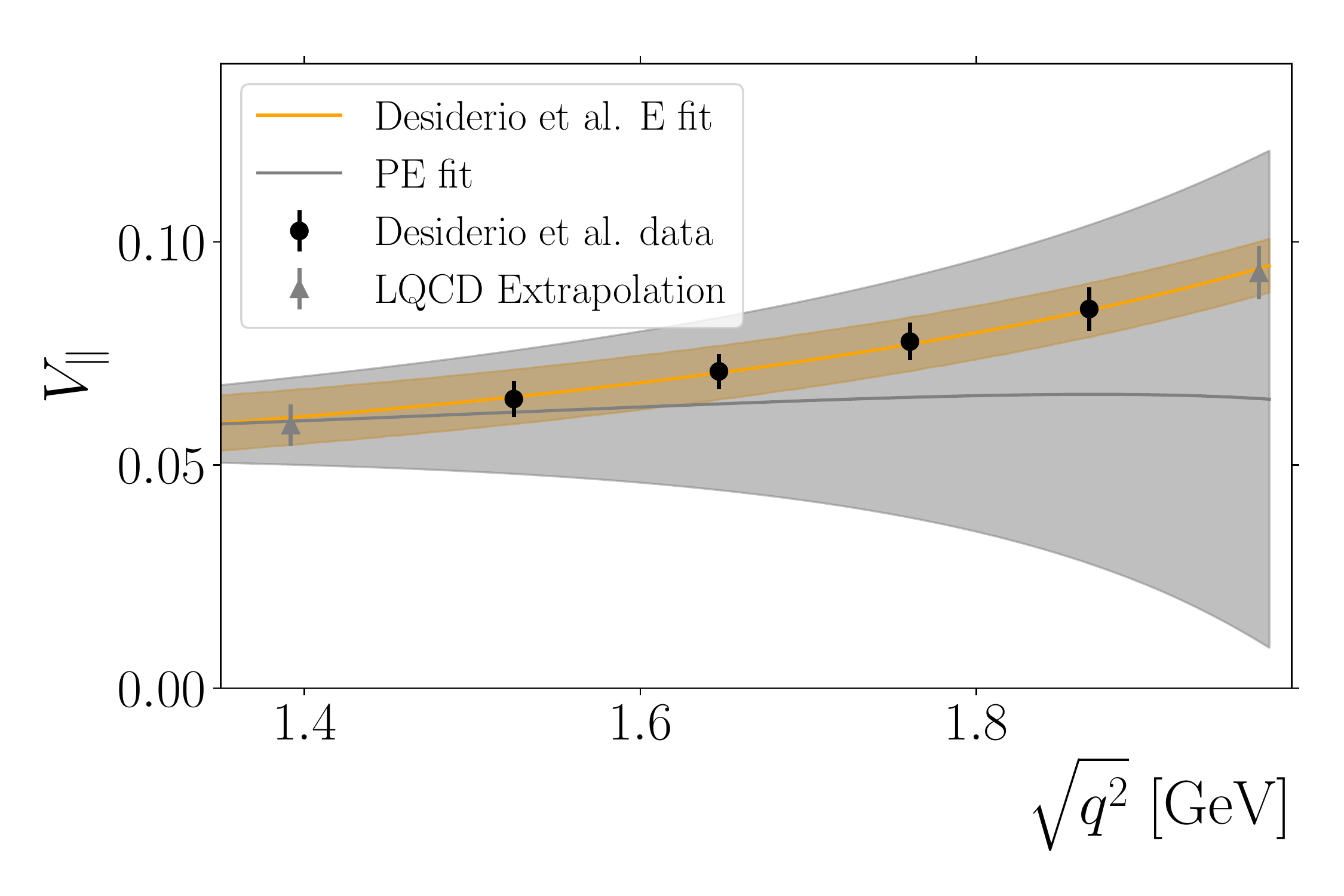}
 \centering
\caption{
Form factor $V_{\parallel}^{D_s} (\sqrt{q^2})$ vs. $\sqrt{q^2}$. Colour code and fitting ansaetze as in fig.~\ref{fig:FV_fits}.}
\label{fig:FA_fits}
\end{figure}

A similar strategy as the one discussed in Sec. \ref{sec:appl_FV} can be applied to $V_{\parallel}^{D_s}(q^2)$ as well. The spectrum of physical resonances in this channel consists of two very closely spaced states, the $D_{s1}(2460)$ and the $D_{s1}(2536)$. We thus expect that a simple {\tt P} ansatz (see eq.~ (\ref{eq:FF=disprel})), with a mass fixed to 2.5 GeV, will fit well the data. We obtain
\be
\label{eq:FA_Pfit}
r_{\parallel 1} = -0.036 (2)~,~~~~ \chi^2 = 13~,~~~~ \mbox{$p$-value}=0.004~,
\ee
corresponding to the blue band in the top left panel of fig.\,\ref{fig:FA_fits}.\footnote{To ease comparison with other results present in literature for the $B_s$ sector, fig.~\ref{fig:FA_fits} shows the absolute values of the FF $V^{D_s}_{\parallel}$ and of LQCD data.} As a cross-check, we also consider an {\tt E} fit, where we let the pole mass be determined by the fit. We get
\bea
\label{eq:FA_Efit}
&r_{\parallel 1} = -0.047 (4)~,~~~~m_{\parallel 1}=2.8 (2)~\GeV~,~~~~\rho({r_{\parallel 1}, m_{\parallel 1}}) = +0.75~,&\nn \\
&\chi^2 = 1.0~.&
\eea
This fit is shown in fig.\,\ref{fig:FA_fits} (top right panel). Comparing the two results we note that, although the physical $D_{s1}(2460)$-meson mass used in the {\tt P} fit is not exactly what is preferred by data, the {\tt P} fit represents a good approximation of the result of the {\tt E} fit, as also clear from fig.\,\ref{fig:FA_fits}. This statement is also supported by the similarity of the residues in the {\tt P} vs. the {\tt E} fit, which is a non-trivial finding---the corresponding comparison fails in the vector channel, for reasons we have also discussed. 

Yet another cross-check is represented by the {\tt PE} fit
\be
\label{eq:FA_PEfit}
V_{\parallel}^{D_s}(x_{\gamma}) = \frac{r_{\parallel 1}}{1 - \frac{m_{D_s}^2(1 - x_{\gamma})}{m_{D_{s1}}^2}} + \frac{r_{\parallel 2}}{1 - \frac{m_{D_s}^2 (1 - x_{\gamma})}{m_{\parallel 2}^2}}~.
\ee
for which we obtain
\bea
&r_{\parallel 1} = 0.03 (7)~,~~~~ r_{\parallel 2} =-0.08 (7)~,~~~~ m_{\parallel 2}=2.9 (4)~\GeV~,&\nn \\
&\rho(r_{\parallel 1},r_{\parallel 2}) =-1~,~~~~ \rho(r_{\parallel 1},m_{\parallel 2}) = +0.63~,~~~~ \rho(r_{\parallel 2},m_{\parallel 2}) = -0.62~,&\nn \\
&\chi^2 = 2.0~.&
\eea
We note in particular that $m_{\parallel 2}$ is consistent with $m_{D_{s1}}$, and that the central value for $r_{\parallel 1} + r_{\parallel 2}$ is very close to the residue determination in either of fits {\tt P} or {\tt E}, eqs. (\ref{eq:FA_Pfit})-(\ref{eq:FA_Efit}), as one may also expect given the proximity of $m_{D_{s1}(2460)}$ and $m_{D_{s1}(2536)}$.
The \PE fit is shown in fig.\,\ref{fig:FA_fits}. As in the vector case, the larger uncertainties affecting both the residues, and especially the mass of the effective pole, translate into a large uncertainty in the predicted FF at high $q^2$.
 
The above results show again a coherent picture, and we take the {\tt P}-fit result as our reference for the axial case. Using eq.~(\ref{eq:res=g}) and the numerical values of the meson decay constants in Appendix \ref{app:meson_decay_constants} we infer the value of the tri-coupling $g_{D_{s1} D_s \gamma}$ as
\be
\label{eq:g_from_P_in_FA}
g_{D_{s1} D_s \gamma} = -0.23 (2) ~{\rm GeV}^{-1}~.
\ee

\medskip

In great synthesis, the different fitting ansaetze attempted do support a well-defined VMD interpretation for both the vector and the axial FF, in the form of a \PP fit in the vector channel and of a {\tt P} fit in the axial one. These results come with a clear-cut advantage for our purposes. Given that the poles in a {\tt P} or \PP fit are physical, these fits lend themselves to an extrapolation procedure guided by heavy-quark scaling, that will allow us to predict the $B_s$ counterparts of these FFs. This is the subject of the next section.

\section{\boldmath Extrapolation from the $D_s$ to the $B_s$ case}
\label{sec:FF_extrapolation}

\subsection{Preliminaries}

Our aim in this section is to use the analysis in Sec. \ref{sec:Ds_FFs} to infer the hadronic FFs entering in $B_s \to \gamma \ell^+ \ell^-$ decays. 

The very first observation to be made is that the $D_s$ and the $B_s$ mesons have different electric charges. In particular, as spelled out in Ref.\,\cite{Desiderio:2020oej}, $V_{\parallel}$ includes a structure-{\em in}dependent contribution proportional to the e.m. charge of the decaying meson and which would dominate at low values of $x_{\gamma}$. Hence this contribution has to be subtracted off the $D_s$-case FF before any extrapolation to the $B_s$ case, where such contribution is absent. Fortunately, the LQCD data in Ref.~\cite{Desiderio:2020oej} are free from this infrared-divergent contribution. Presenting FFs with such contribution subtracted, as advocated in Ref.~\cite{Desiderio:2020oej}, is thus advantageous at least in our context.

Our $D_s$-case analysis in Sec.~\ref{sec:Ds_FFs} led to the conclusion that a {\tt P} fit is adequate to describe the axial-vector FF $V_{\parallel}$, whereas we necessitate a {\tt PP} ansatz for the vector FF $V_{\perp}$. We assume the same functional forms in the $B_s$ case. We accordingly describe the FF $V_{\parallel}^{B_s}(q^2)$ through a {\tt P} ansatz, and the FF $V_{\perp}^{B_s}(q^2)$ through a \PP ansatz. Needless to say, this assumption can only be validated once the ``$B_s$ counterpart'' of Ref.~\cite{Desiderio:2020oej} will be available. In the axial-vector case, the mass of the {\tt P}-pole particle, the $B_{s1}(5830)$, is taken from the PDG \cite{ParticleDataGroup:2022pth}; in the vector case, the first mass required by the \PP fit, the $B_s^*$'s, is again taken from the PDG. We note at this juncture that, although heavier resonances have been observed, their $J^P$ quantum numbers are yet to be established. As a consequence, our second pole mass required, to be referred to as $B_{s1}^*$ in analogy with the $D_s$-case, is taken from a quark model, see the $2^3 S_1$ row of table 1 in Ref.~\cite{Godfrey:2016nwn}, $m_{B_{s1}^*} = 6.012(50) ~{\rm GeV}$. With these ingredients, we can thus write
\bea
\label{eq:FVBs}
V_{\perp}^{B_s} (q^2) &=& \frac{r_{\perp 1}^{B_s}}{1-q^2/m_{B_s^*}^2} + \frac{r_{\perp 2}^{B_s}}{1-q^2/m_{B_{s1}^*}^2}~,\\
\label{eq:FABs}
V_{\parallel}^{B_s} (q^2) &=& \frac{r_{\parallel}^{B_s}}{1-q^2/m_{B_{s1}}^2}~.
\eea
The residues $r_{\perp 1}^{B_s}$, $r_{\perp 2}^{B_s}$ and $r_{\parallel}^{B_s}$ are expected to obey relations analogous to the ones in eq.\,(\ref{eq:res=g}). Their $D_s$-sector counterparts have been determined in Sec.~\ref{sec:Ds_FFs} through direct fits of LQCD data, see in particular eqs. (\ref{eq:FV_PPfit}) and (\ref{eq:FA_Pfit}). We want to use these determinations as a starting point for an extrapolation to the $B_s$ sector, that we describe next.

\subsection{Parameterizing the tri-couplings with the quarks' magnetic moments}\label{sec:g=QxMM}

The residues in eqs. (\ref{eq:FVBs}) and (\ref{eq:FABs}) can be related to $g_{B_{sJ}^{(*)} B_s \gamma}$ tri-couplings (with $B_{sJ}^{(*)}$ denoting the appropriate excited state in the vector or axial channel) via relations similar to eqs.~(\ref{eq:res=g}). Following Ref.~\cite{Becirevic:2009aq}, these tri-couplings can in turn be parameterized as the sum of the magnetic moments of the valence quarks, each term in the sum being weighted with the electric charge of the corresponding quark. At variance with Ref.~\cite{Becirevic:2009aq}, we consider both the heavy- and the light-quark terms in such sums, i.e. we do not sit in the heavy-quark limit. Hence, in the axial case this parameterization yields
\bea
\label{eq:gA=sumMMxQ}
g_{D_{s1}D_s\gamma} &=& - Q_s \mu^\parallel_s + Q_c \mu^\parallel_c~, \nn \\
g_{B_{s1}B_s\gamma} &=& - Q_s \mu^\parallel_s + Q_b \mu^\parallel_b~, 
\eea
where the quantities $\mu^\parallel_s$, $\mu^\parallel_c$ snd $\mu^\parallel_b$ are the strange-, charm-, and bottom-quark magnetic moments in the axial channel and, obviously, $Q_s = Q_b = -1/3$, $Q_c = +2/3$.

The above parameterization is useful because the magnetic moment of a fermion scales with the inverse of the mass of the same fermion. Then we may express $\mu^\parallel_c$ and $\mu^\parallel_b$ as functions of $\mu^\parallel_s$, i.e.
\be
\label{eq:MM_h_vs_l}
\mu^\parallel_c = \frac{m_s}{m_c} \mu^\parallel_s~,~~~~ \mu^\parallel_b = \frac{m_s}{m_b} \mu^\parallel_s~.
\ee
In these relations, the quark masses are to be understood as ``constituent'' masses, with values to be taken from e.g. Refs.\,\cite{Godfrey:2016nwn, Godfrey:2015dva}. In such approach, the only unknown quantity describing both $g_{D_{s1}D_s\gamma}$ and $g_{B_{s1}B_s\gamma}$ is thus $\mu^\parallel_s$ that can, however, be directly inferred from the results of Sec.~\ref{sec:Ds_FFs}. Hence, taking the values of the meson decay constants $f_{D_{s1}}$ and $f_{B_{s1}}$ from Appendix \ref{app:meson_decay_constants} and recalling eqs.\,(\ref{eq:res=g}) and (\ref{eq:FABs})-(\ref{eq:gA=sumMMxQ}), we obtain a determination of the axial FF $V_{\parallel}^{B_s}$. 

This strategy can be extended to the vector FF $V_\perp^{B_s}$. Our starting point in this case, $V_\perp^{D_s}$, has been found to require {\em two} poles, see Sec. \ref{sec:appl_FV}. This implies two different tri-couplings in the $D_s$ sector, that we want to extrapolate to the $B_s$ sector. We parameterize these couplings, as described above, in terms of magnetic moments times charges of the respective quarks\footnote{Note that similar relations can be found for instance in \cite{Donald:2013sra} for the study of the $D_s^* \to D_s \gamma$ decays on the lattice.}   
\bea
\label{eq:gV=sumMMxQ}
g_{D_{s}^*D_s\gamma} &=& Q_s \mu^{\perp 1}_s + Q_c \mu^{\perp 1}_c~, \nn \\
g_{D_{s1}^*D_s\gamma} &=& Q_s \mu^{\perp 2}_{s} + Q_c \mu^{\perp 2}_{c}~, \nn \\
g_{B_{s}^*B_s\gamma} &=& Q_s \mu^{\perp 1}_s + Q_b \mu^{\perp 1}_b~, \nonumber \\
g_{B_{s1}^*B_s\gamma} &=& Q_s \mu^{\perp 2}_{s} + Q_b \mu^{\perp 2}_{b}~.
\eea
Note that we have introduced two different magnetic moments, for instance $\mu_s^{\perp 1}$ and $\mu_{s}^{\perp 2}$ in the case of the strange quark, which allow to distinguish between the two different poles.\footnote{The sign difference in the strange-quark contribution in eqs.\,(\ref{eq:gA=sumMMxQ}) and (\ref{eq:gV=sumMMxQ}) is due to the opposite behaviour of the vector and the axial currents under charge conjugation. In the limit of degenerate valence quarks, this sign difference implies that the axial-sector tri-coupling vanishes. This is supported by the decay pattern of, for instance, $c\bar{c}$ states. While the vector-$c \bar c$ decay $J/\Psi \to \eta_c \gamma$ exists, the axial-$c \bar c$ counterpart $\chi_{1c} \to \eta_c \gamma$ does not~\cite{ParticleDataGroup:2022pth}.}

We then determine the magnetic moment $\mu^{\perp i}_s$ from the residue $r_{\perp i}^{D_s}$ ($i=1,2$) in eq.\,(\ref{eq:FV_PPfit}), taking into account the correlation between the two residues, $\rho(r_{\perp 1}, r_{\perp 2}) = -0.44$ as in eq.\,(\ref{eq:FV_PPfit}). To this end, we performed a ``bootstrap'' analysis, where we sample $10^5$ instances of the residues $r_{\perp 1}^{D_s}$ and $r_{\perp 2}^{D_s}$ through a multivariate Gaussian distribution. From this sample we then infer mean values, uncertainties and correlation of the $B_s$-sector counterparts, $r_{\perp 1}^{B_s}$ and $r_{\perp 2}^{B_s}$.

\begin{table}[h!]
\renewcommand{\arraystretch}{1.5}
\begin{center}
\begin{tabular}{|c|c||c|c|}
\hline
\multicolumn{2}{|c||}{\gcl Quark magnetic moments} & \multicolumn{2}{c|}{\gcl $B_s \to \gamma$ FFs parameters } \\
\hline
\hline
$\mu_s^{\perp 1}$  & $-0.22 (8)$   & $r_{\perp 1}^{B_s}$             & $0.017 \pm 0.006$ \\ 
$\mu_b^{\perp 1}$  & $-0.019 (6)$  & $r_{\perp 2}^{B_s}$             & $0.088 \pm 0.030$ \\
$\mu_s^{\perp 2}$  & $-2.6 (8)$    & $r_{\parallel}^{B_s}$               &  $-0.043 \pm 0.004$ \\
$\mu_b^{\perp 2}$  & $-0.22 (6)$   & $\rho(r_{\perp 1},r_{\perp 2})$  & $-0.21$            \\           
$\mu_s^{\parallel}$ & $-0.46 (4)$  & $ $ & $ $ \\ 
$\mu_b^{\parallel}$ & $-0.038 (3)$ & $ $ & $ $  \\
\hline
\end{tabular}
\end{center}
\renewcommand{\arraystretch}{1.0}
\caption{(Left) Numerical values for the valence-quarks' magnetic moments in the vector and axial sectors, as inferred from LQCD data \cite{Desiderio:2020oej} and from the scaling in eq.\,(\ref{eq:MM_h_vs_l}). (Right) Parameters of the \Bstog FFs following the notations of eqs. (\ref{eq:FVBs}), (\ref{eq:FABs}).}
\label{tab:mag_mom}
\end{table}
\subsection{\boldmath Results for the \FVA form factors} \label{sec:results_Bs_FFs}

\begin{figure}[h!]
 \centering
\includegraphics[width=0.49\textwidth]{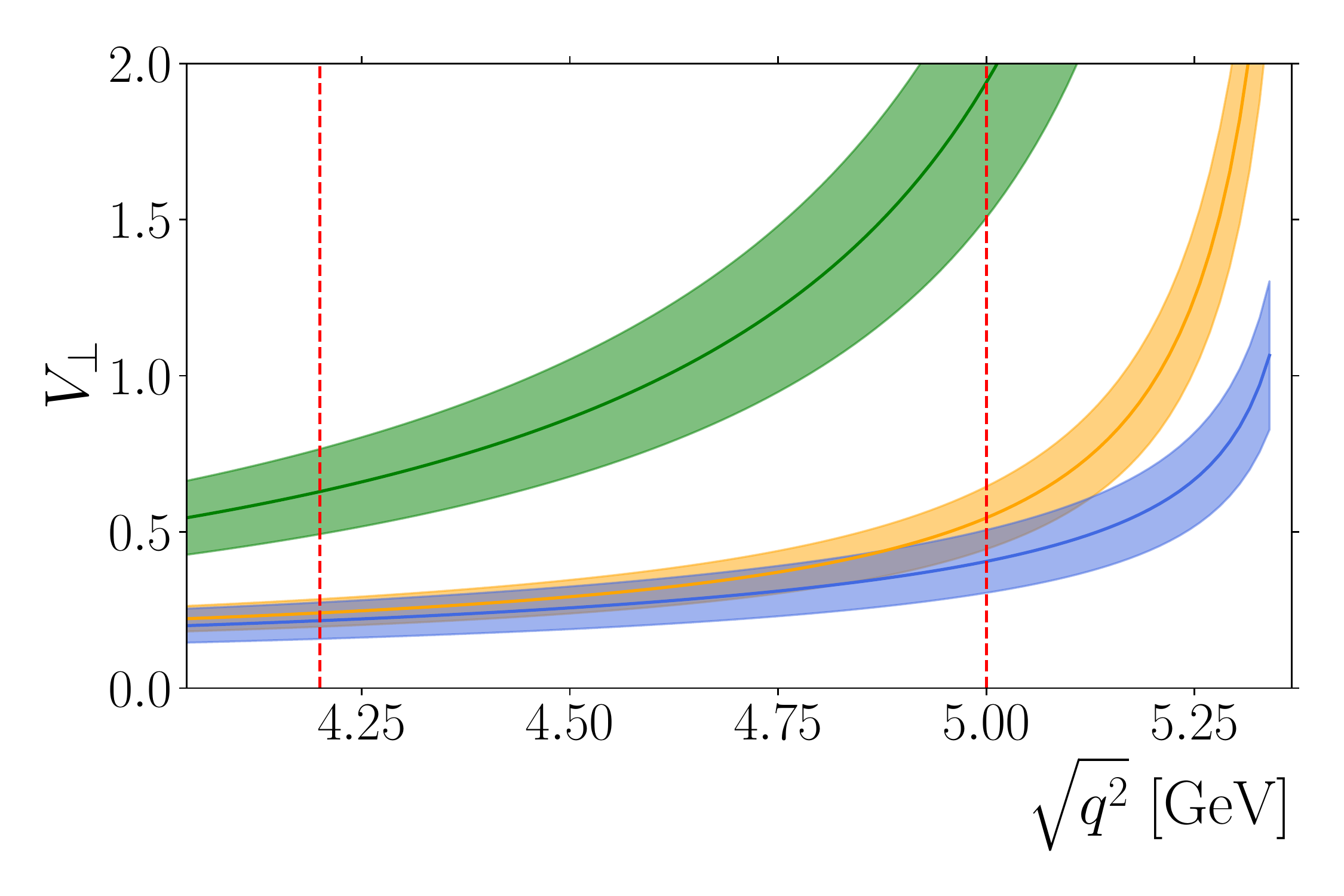}
\includegraphics[width=0.49\textwidth]{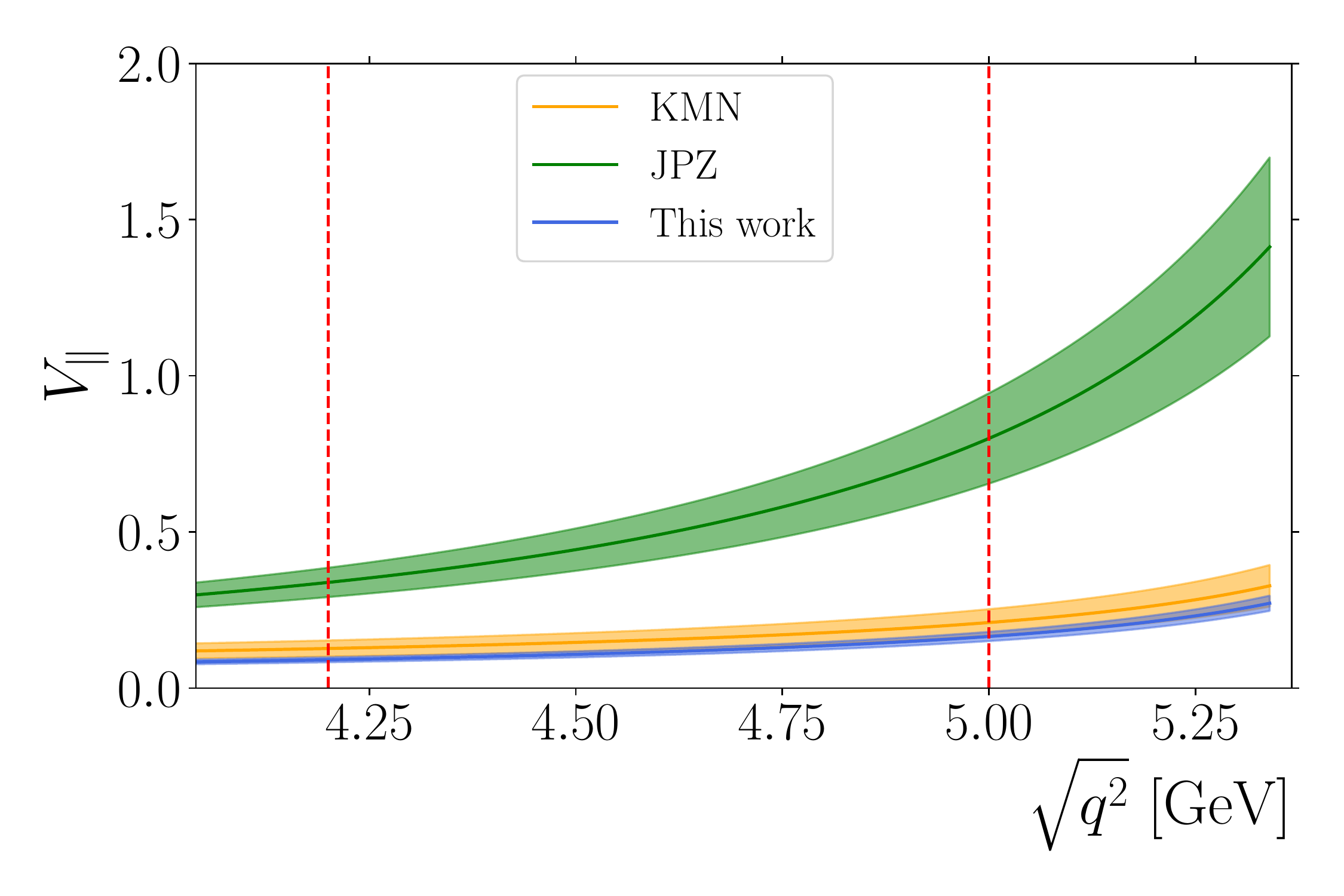}
 \centering
\caption{Form factors $V_{\perp}^{B_s} (q^2)$ (left panel) and $V_{\parallel}^{B_s} (q^2)$ (right panel). The colour code is detailed in the legend. The acronyms KMN and JPZ refer to the determinations in Refs. \cite{Kozachuk:2017mdk} and \cite{Janowski:2021yvz}. The vertical lines at $4.2$ and $5.0$ GeV mark a possible reference region for the ``indirect'' measurement of $\mc B(B_s \to \mu^+ \mu^- \gamma)$.}
\label{fig:FFsBs}
\end{figure}

\noindent This section presents our results for $V_{\perp,\parallel}^{B_s}$ as obtained from the parameterization and extrapolation procedure described above. Table \ref{tab:mag_mom} shows the values of the vector and the axial magnetic moments of the strange quark and the bottom quark, and lists the corresponding residues of the bottom sector, that enter the parameterization in eqs.~(\ref{eq:FVBs}), (\ref{eq:FABs}). Figure~\ref{fig:FFsBs} displays the corresponding FFs entering the $B_s \to \gamma$ decay, namely $V_{\perp}^{B_s}$ (left panel) and $V_{\parallel}^{B_s}$ (right panel).\footnote{In the right panel of fig.~\ref{fig:FFsBs}, we plot the absolute value of $V_{\parallel}^{B_s}$, as done for $V_{\parallel}^{D_s}$.} Our determination is compared with the other ones in literature mentioned in the Introduction, namely the quark-model FFs of Ref. \cite{Kozachuk:2017mdk} updating Ref. \cite{Melikhov:2004mk}, and the LCSR FFs of Ref. \cite{Janowski:2021yvz}. The main comment of practical importance to be made is that, in the indicative range $\sqrt{q^2} \in [4.2, 5.0]$ GeV relevant for the ``indirect'' measurement \cite{Dettori:2016zff} of $\mc B(B_s \to \mu^+ \mu^- \gamma)$, our determination is in good agreement with the KMN one, while being largely inconsistent with the JPZ one. Specifically, the figure suggests a difference in $V_\perp$ (the FF giving the  dominant contribution in our $q^2$ range of interest) by a factor between $\sim 2$ and $\sim 5$ at respectively the lower and upper bounds of the mentioned $\sqrt{q^2}$ range. This has strong consequences on the prediction of the integrated branching ratio, recalling that the latter has quadratic dependence on the FFs. This difference may be traced back to the different values of the tri-couplings.

\subsection{On the heavy-quark scaling of the meson decay constants}

In principle, another possibility to infer the behaviour of the hadronic FFs in $B_s \to \ell^+ \ell^- \gamma$ decays would be to assume a well-defined scaling of the meson decay constants $f$ in the heavy-quark (HQ) limit. A scaling often quoted in the literature is $f \sim m_q^{-1/2}$, where $m_q$ is the mass of the HQ. Then, starting from the $D_s$-sector decay constants, we may infer the $B_s$-sector ones (or viceversa), using this scaling. 

In the vector-FF case, we observe general consistency between the determination shown in blue in fig. \ref{fig:FFsBs}---that does not make any scaling assumption on the decay constants---and the determination obtained by scaling the decay constants as described. On the other hand, we observe sizeable differences in the axial-FF case. The reason for the difference is to be traced back in the numerical values of the meson decay constants listed in App.~\ref{app:meson_decay_constants}. In fact, by using those values we can determine the scaling of the meson decay constants directly from LQCD data, separately in the vector and axial channel. Denoting the resonances as $B_{V_1,V_2,A}$ and $D_{V_1,V_2, A}$, this means that
\be
\label{eq:f_scaling}
f_{B_{V_1,V_2,A}} (m_{B_{V_1,V_2,A}})^{n_{V_1,V_2,A}} = f_{D_{V_1,V_2,A}} (m_{D_{V_1,V_2,A}})^{n_{V_1,V_2,A}}~,
\ee
neglecting short-distance corrections \cite{Neubert:1992fk}. In the pseudoscalar-meson case $n_{V_1,V_2,A}$ is assumed to be $1/2$. Note that eq. (\ref{eq:f_scaling}) allows the two poles in the $1^-$ channel to have different scaling relations. Eq. (\ref{eq:f_scaling}) implies
\be
n_{V_1,V_2,A} = \frac{\log(f_{B_{V_1,V_2,A}}/f_{D_{V_1,V_2,A}})}{\log(m_{D_{V_1,V_2,A}}/m_{B_{V_1,V_2,A}})}~.
\ee
Taking the parameters on the r.h.s. from the data in App. \ref{app:meson_decay_constants} we find
\be
\label{eq:nV1V2A}
n_{V_1} = 0.212 (22)~,~~~~ n_{V_2} = -0.07 (25)~,~~~~ n_A = -0.630 (89)~,
\ee
where we have simply propagated the uncertainties affecting the meson decay constants. These relations, and especially the axial-channel ones, display a sizeable departure from the usual HQ scaling $n \simeq 1/2$.

This discussion warrants further investigations of the coefficients $n_{V_1,V_2,A}$ directly on the lattice. From a practical point of view, and as already mentioned, sizeable differences in the axial channel (as opposed to the vector channel) have only a limited impact on the SM prediction for $\mc B(B_s \to \mu^+ \mu^- \gamma)$ at high $q^2$, to which we turn next.

\section{Prediction of the \boldBsmmy branching fraction}
\label{sec:Bsmumuy_prediction}

\subsection{Preliminaries}\label{sec:preliminaries}

Using our results on the \Bstog FFs, we now provide a SM prediction of the single-differential (in $q^2$) as well as integrated branching fraction of the \Bsmmy decay, in the region of high $q^2$. The \Bsmmy amplitude consists of two components, one known as ``direct emission'' (DE), $\mc A_{\DE}$, where both the weak operator and the e.m. current have to be evaluated between the external meson and photon; the other known as bremsstrahlung, $\mc A_{\Brems}$, where the e.m. current is evaluated between the external di-lepton and the vacuum. Explicit formulae may be found in existing literature and will not be repeated here. Specifically, for notation and a clearheaded overall discussion we refer the reader to Ref.~\cite{Beneke:2020fot}; useful formulae can be found in Refs.~\cite{Melikhov:2004mk,Kozachuk:2017mdk,Guadagnoli:2016erb}---the latter discussing the sign of the interference term according to the $f_B$ convention followed.

The amplitude's calculation involves four hadronic matrix elements. The vector and axial ones, whose FFs have been the focus of the preceding sections, have been defined in eq.\,(\ref{eq:FV_FA_defs}). The two further matrix elements necessary are the tensor and axial-tensor ones, defined as
\bea
\label{eq:FTV_FTA_defs}
\< \gamma(k,\la)|\bar s \sigma^{\mu \nu} b q_\nu| \bBs(q+k) \> &=& 
i e \, \eps^{\mu \la^* q k} \FTV(q^2,0)\,, \nn \\
[0.2cm]
\< \gamma(k,\la)|\bar s \sigma^{\mu \nu} \gamma_5 b q_\nu| \bBs(q+k) \> &=& 
e \, (\la^{*\mu} \, q k - k^\mu \, \la^* q) \FTA(q^2,0)\,.
\eea
with the shorthand $\eps^{\mu \la^* q k} \equiv \eps^{\mu \alpha \beta \delta}\la^{*\alpha} q^{\beta} k^{\delta}$ and the notation dictionary $F_{TV} = - T_\perp$, $F_{TA} = -T_\parallel$, see e.g. \cite{Guadagnoli:2017quo}.

With the \Bsmmy amplitude one can compute the single-differential branching fraction in $q^2$ as a sum of three components, often denoted as $d \Gamma^{(1), (2), (12)} / dq^2$ and due to DE, to bremsstrahlung, and to the interference between the two, respectively. The interference component is negligible throughout the full kinematic range;\footnote{We find variations $\lesssim 3\%$ in the integrated observable in our range of interest, when calculating it with $\Gamma^{(1)}$ alone or with $\Gamma^{(1)} + \Gamma^{(12)}$. Such variation is well within the current theoretical error, dominated by the FF determination. This variation should however be kept in mind, and can trivially be taken into account if the FF error were to shrink to the percent level.} the DE component quickly dominates for $\sqrt{q^2}$ below $5.0~\GeV$ \cite{Dettori:2016zff}; finally, the bremsstrahlung component is summed to all orders and corrected for in the \Bsmm observable \cite{Davidson:2010ew}. These circumstances make the DE-only component of $\mc B(\Bsmmy)$ a well-defined observable in the range $\sqrt{q^2} \in [4.2, 5.0]~\GeV$. This component coincides with the $\Bsmmy$ contribution fitted in the analyses of Refs. \cite{LHCb:2021vsc,LHCb:2021awg} along with the purely leptonic modes, the latter understood to be fully photon-inclusive \cite{Buras:2012ru}. Besides, all $B_s$ decays are understood to be corrected for the sizeable lifetime difference of the mass eigenstates of the $B_s^0 - \bar{B}_s^0$ system \cite{Dunietz:2000cr,Descotes-Genon:2011rgs,DeBruyn:2012wj}.
We next discuss the inclusion and treatment of the different sources of hadronic uncertainties.

\subsection{Form-factor parametrization}

For the $B_s$-sector \FVA FFs we use the parameterization summarized in table \ref{tab:mag_mom}, and discussed in Secs. \ref{sec:Ds_FFs}-\ref{sec:FF_extrapolation}. On the other hand, our approach does not give us access to the tensor FFs $T_{\perp, \parallel}$, see eq. (\ref{eq:FTV_FTA_defs}).
In the absence of a lattice-QCD computation of these quantities in the high-$q^2$ region, whether in the $D_s$ sector or directly in the $B_s$ one, we estimate their impact on the prediction of our observable of interest by either resorting to the KMN or JPZ determinations in \cite{Kozachuk:2017mdk,Janowski:2021yvz}, or else by setting $T_{\perp, \parallel} = 0$. This approximation is meaningful because $T_{\perp, \parallel}$ give small contributions in the high-$q^2$ region of concern to us. When adopting the JPZ parametrization, we accordingly use the standard deviations and correlations it comes with; in the case of the KMN parametrization, we consider variations of $\pm 20\%$ around the central values provided for $F_i(0)$, $i=V,A,TV,TA$, see \cite{Kozachuk:2017mdk}. Such $20\%$ figure is not meant as a realistic assessment of the KMN-FF errors, which are simply unknown; we use it for indicative purposes, to namely provide an idea of the impact of a 20\% FF variation on the $\mc B(\Bsmmy)$ prediction.

\subsection{Charmonium resonances}\label{sec:broad_cc}

Our kinematic region of interest, $\sqrt{q^2} \in [4.2, 5.0]~\GeV$, is close to, or it overlaps with, the mass peaks of the broad-charmonium resonances $\psi(3770)$, $\psi(4040)$, $\psi(4160)$, $\psi(4415)$. (In our numerics we also include the $\psi(2S)$, whose mass peak $m_{\psi(2S)} = 3686.10(6)~\MeV$ is narrower and sizeably below our reference kinematic region.)
In order to include these resonances in the amplitude, we follow the approach of Ref. \cite{Kruger:1996cv} where resonances are included as properly normalized Breit-Wigner (BW) poles that shift the Wilson coefficient $C_9$. In our case, such shift involves the five vector mesons $V$ mentioned, and reads
\be
\label{eq:Vcc_shift}
C_9 ~\to~ C_9 ~-~ \frac{9 \pi}{\alpha^2} \, \bar C \,  \sum_V 
|\eta_V| e^{i \delta_V} \frac{\hat{m}_V \, \mc B(V \to \mu^+ \mu^-) \, \hat{\Gamma}_{\textrm{tot}}^V}{\hat{q}^2 - \hat{m}_V^2 + i \hat{m}_V \hat{\Gamma}^V_{\textrm{tot}}}~.
\ee
Here $\bar C = C_1 + C_{2}/3 + C_3 + C_{4}/3 + C_5 + C_{6}/3$, and $C_9$ stands for $C_9^{\rm eff}(q^2)$, the sum total of the perturbatively calculable contributions \cite{Chetyrkin:1996vx,Bobeth:1999mk}. Hatted quantities are normalized by the appropriate power of $m_{\Bs}$ to make them dimensionless. As noted in Ref. \cite{Beneke:2020fot}, the resonant shift above will not lead to a double counting of part of the short-distance contributions, because this shift is (formally) of higher order in the heavy-quark expansion.

The uncertainty inherent in this shift is encoded in the BW normalisation factors and phases that we scan over with uniform and independent distributions in the ranges $|\eta_V| \in [1, 3]$, $\delta_V \in [0, 2\pi)$. This approach is expected to provide a conservative way to account for deviations from naive factorisation ($|\eta_V| = 1$ and $\delta_V = 0$).\footnote{It was found that $|\eta_V| \simeq 2.5$ and $\delta_V \simeq \pi$ well describe $B \to K \mu^+ \mu^-$ data \cite{Lyon:2014hpa,LHCb:2013ywr}.} We take unity as the reference value for all of $|\eta_V|$, whereas for $\delta_V$ we use the central values from the BESIII determination in Ref.~\cite{BES:2007zwq}.

\subsection{Numerical analysis}

We are now in a position to discuss the prediction of the branching fraction of \Bsmmy in the high-$q^2$ region where it can be measured through the indirect method. The numerical inputs other than the FF parametrization in table~\ref{tab:mag_mom} are summarized in App.~ \ref{app:input_table}. We present our results in two different forms: on the one hand the single-differential branching fraction in $q^2$, on the other the integrated branching fraction in the high-$q^2$ region as a function of the lower bound of integration $q^2_\text{min}$, defined as
\be \label{eq:int-BR}
\mc B(\bsmumugamma)[\sqrt{q^2}_\text{min}, m_{\Bs}] = \int_{q^2_\text{min}}^{m_{\Bs}^2} \frac{d\mc B}{dq^2}dq^2 ~.
\ee
This definition allows to directly compare with the experimental measurement of the ISR component, e.g. the upper limit set by LHCb~\cite{LHCb:2021vsc,LHCb:2021awg} uses $q^2_\text{min} = (\SI{4.9}{GeV})^2$.

\begin{figure}[h!]
\centering
\includegraphics[width=0.49\textwidth]{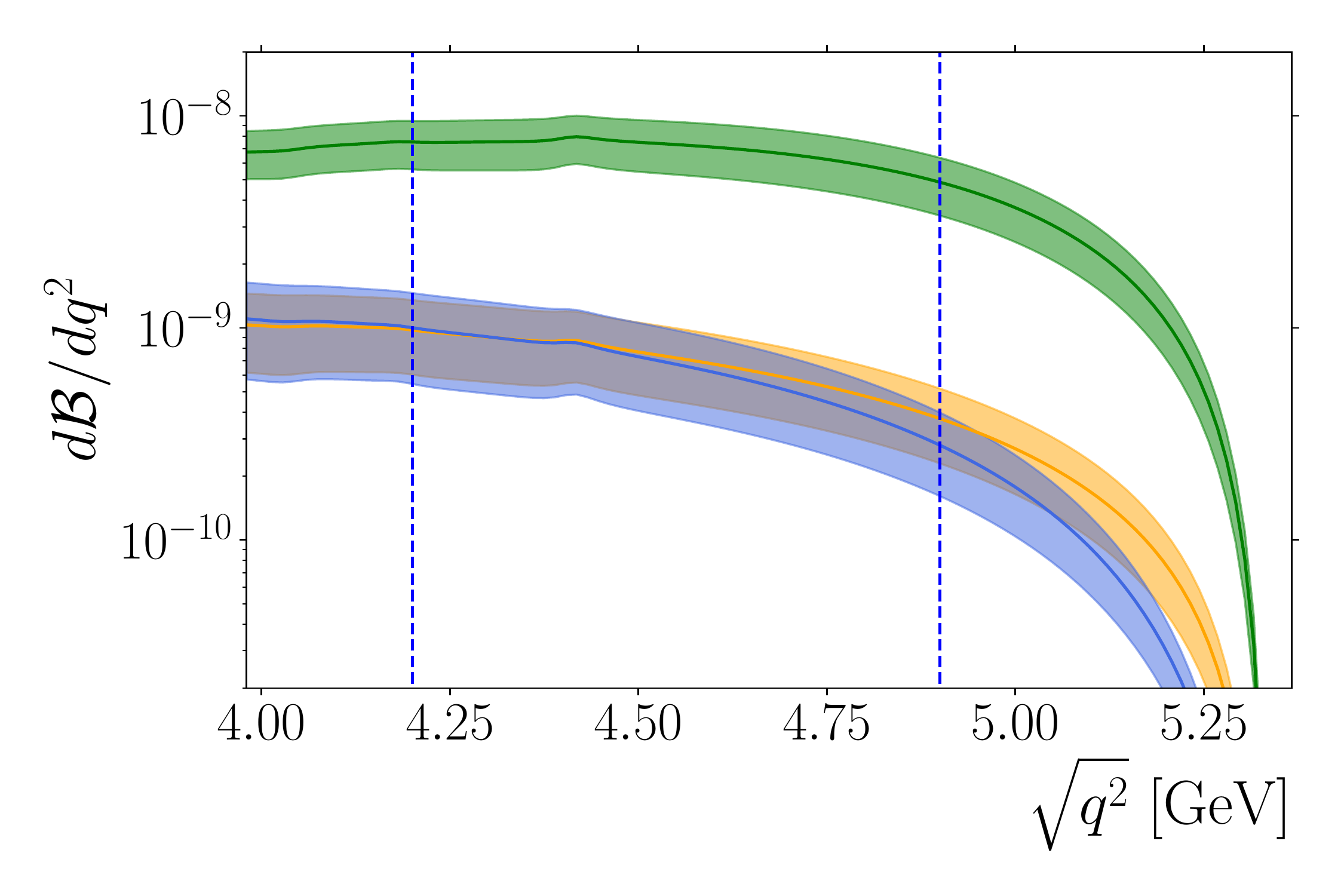}
\includegraphics[width=0.49\textwidth]{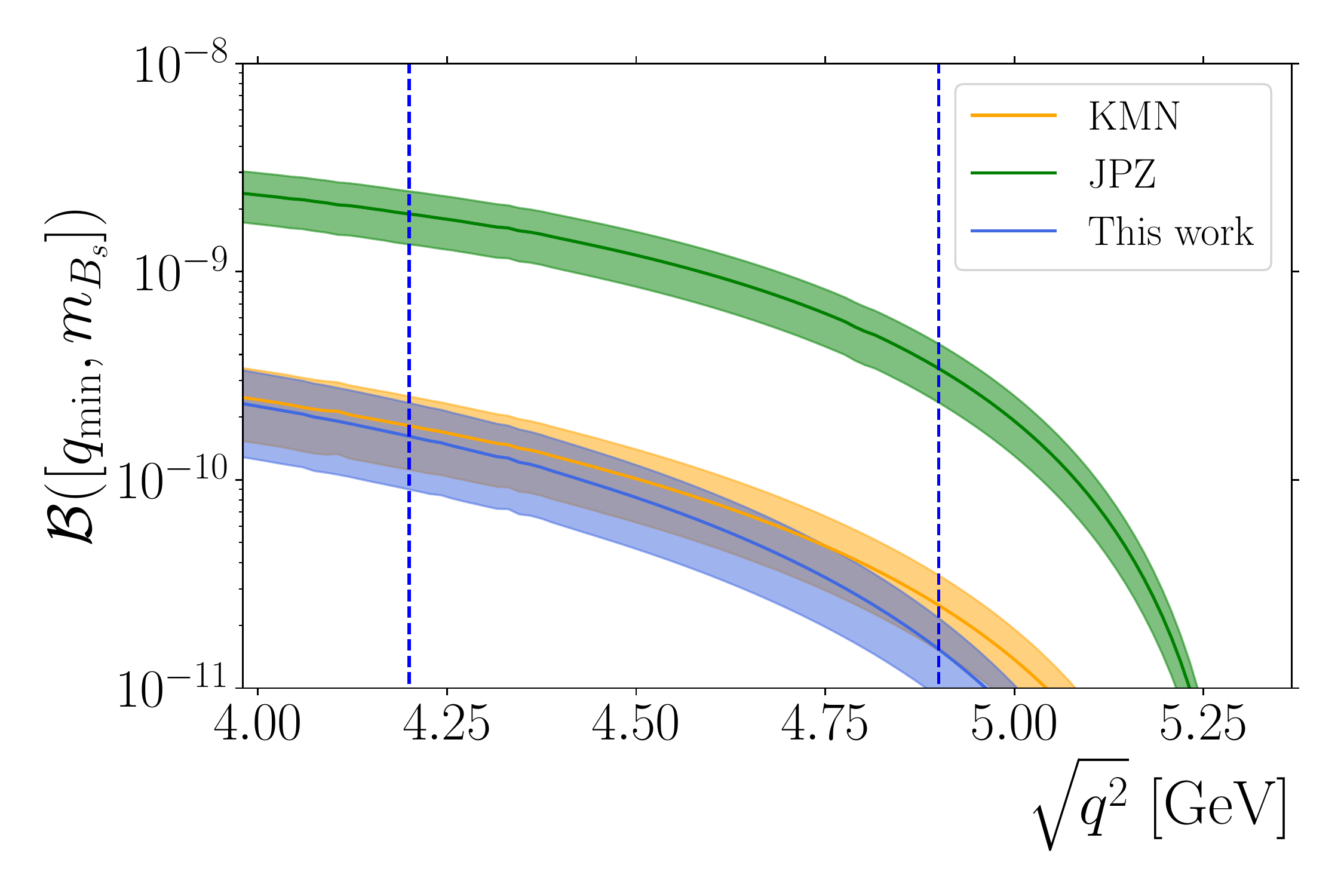}
\centering
\caption{(Left) $d \mc B(\Bsmmy) d q^2$ using the form-factor parametrization from this work (blue), Ref.~\cite{Kozachuk:2017mdk} (orange) or Ref.~\cite{Janowski:2021yvz} (green), see also color code in the legend. (Right) Integrated branching fraction as a function of $\sqrt{q^2}_\text{min}$ as defined in eq.\,(\ref{eq:int-BR}). The vertical dashed lines correspond to different $q^2_{\rm min}$-values for the integration range, in particular $q^2 = (4.9\,\GeV)^2$ has been used by the LHCb analysis of Refs.~\cite{LHCb:2021vsc,LHCb:2021awg}; the value $q^2  = (4.2\,\GeV)^2$ represents a realistic lower limit for a more extended analysis.}
\label{fig:KMN-vs-JPZ-vs-GNSV}
\end{figure}

The first result of importance is the difference in the prediction obtained with the three parametrizations under consideration in this work. In fig. \ref{fig:KMN-vs-JPZ-vs-GNSV}, we compare the ISR component of eq.~(\ref{eq:int-BR}) in the range $[4.2~\GeV,m_{\Bs}]$, with all sources of uncertainties taken into account. As expected, our parametrization is well below---by an order of magnitude---the prediction using the LCSR computation from Ref. \cite{Janowski:2021yvz}, and turns out to be in good agreement with the computation of Ref.~\cite{Kozachuk:2017mdk}.

\begin{figure}[h!]
\centering
\includegraphics[width=0.49\textwidth]{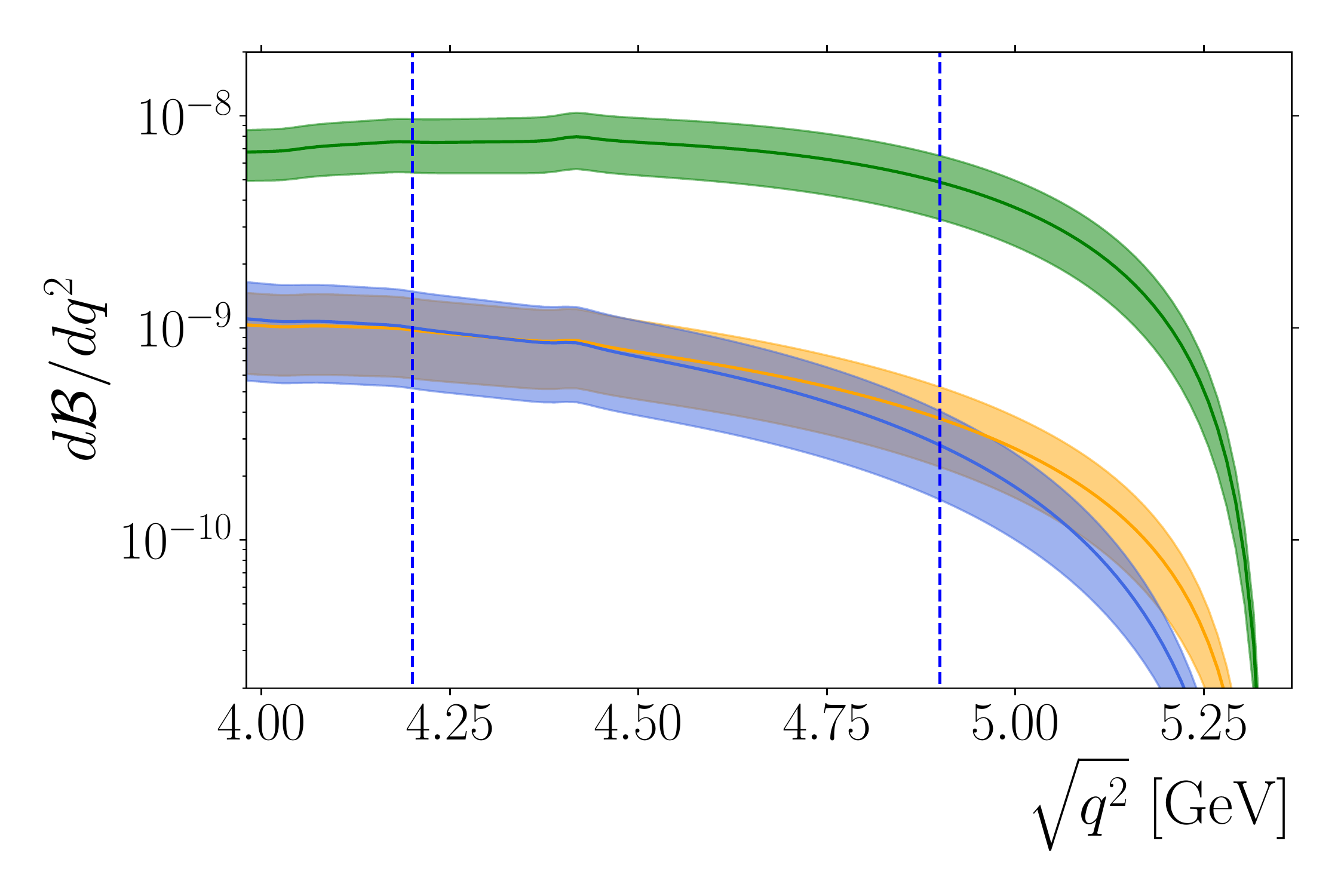}
\includegraphics[width=0.49\textwidth]{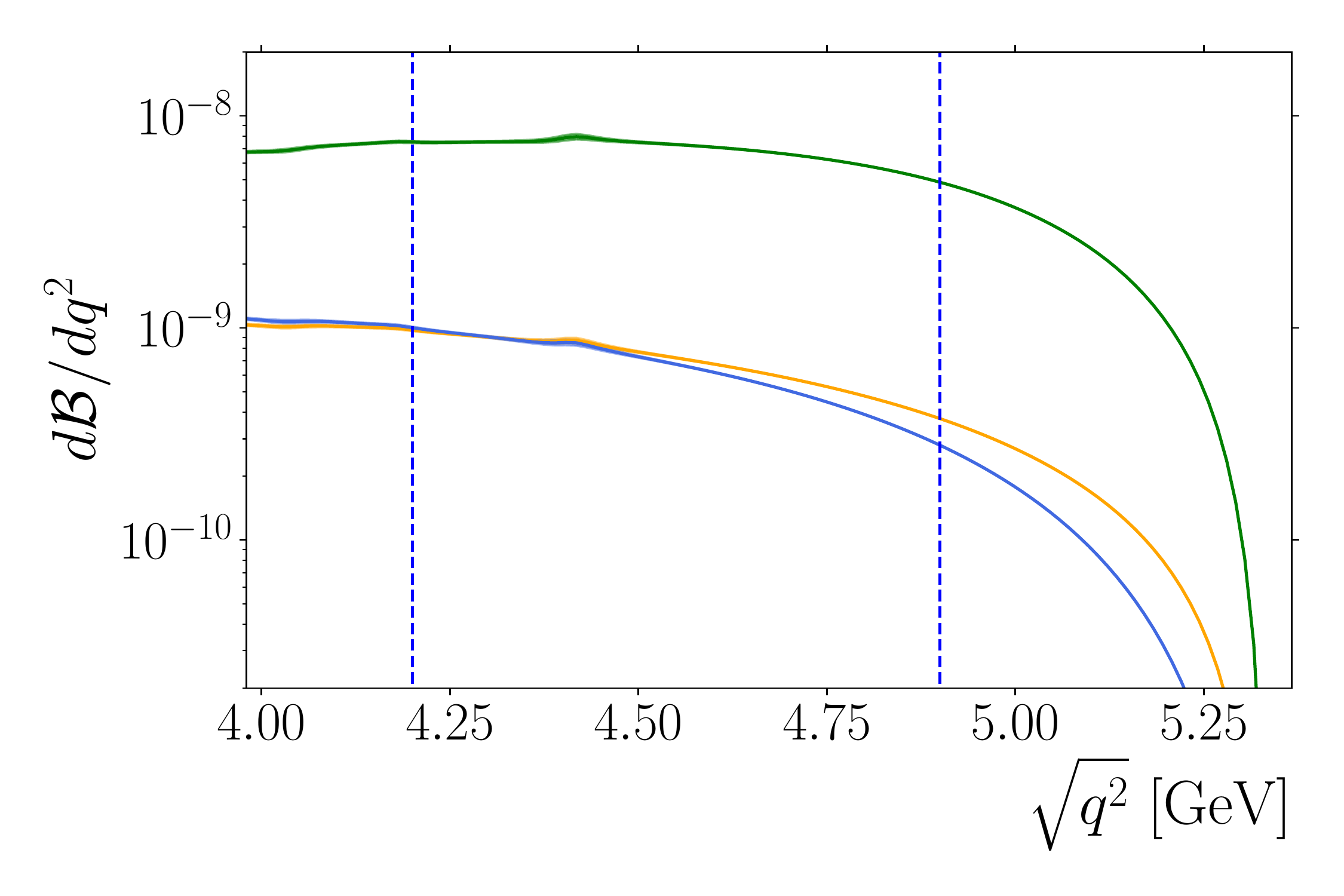}
\includegraphics[width=0.49\textwidth]{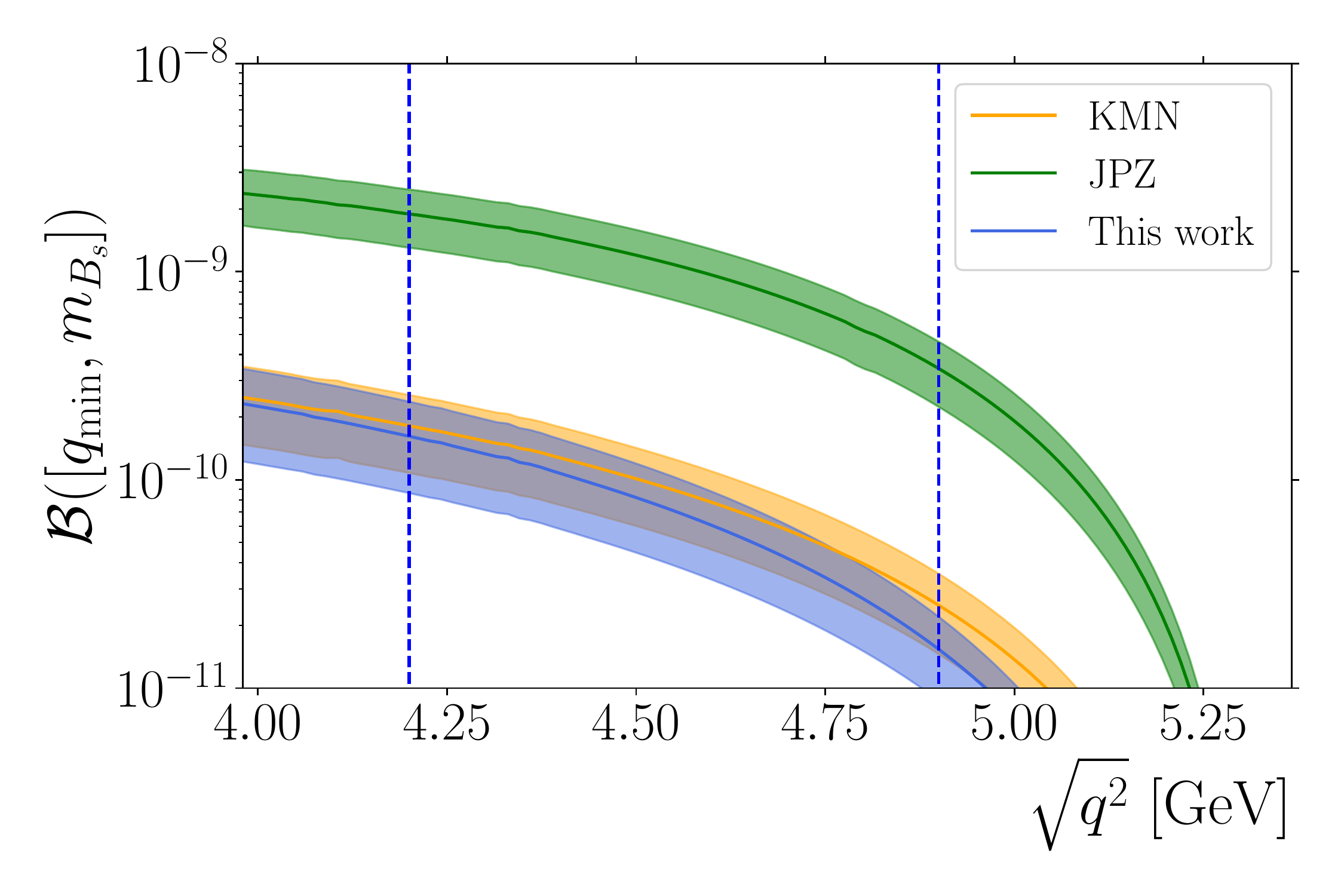}
\includegraphics[width=0.49\textwidth]{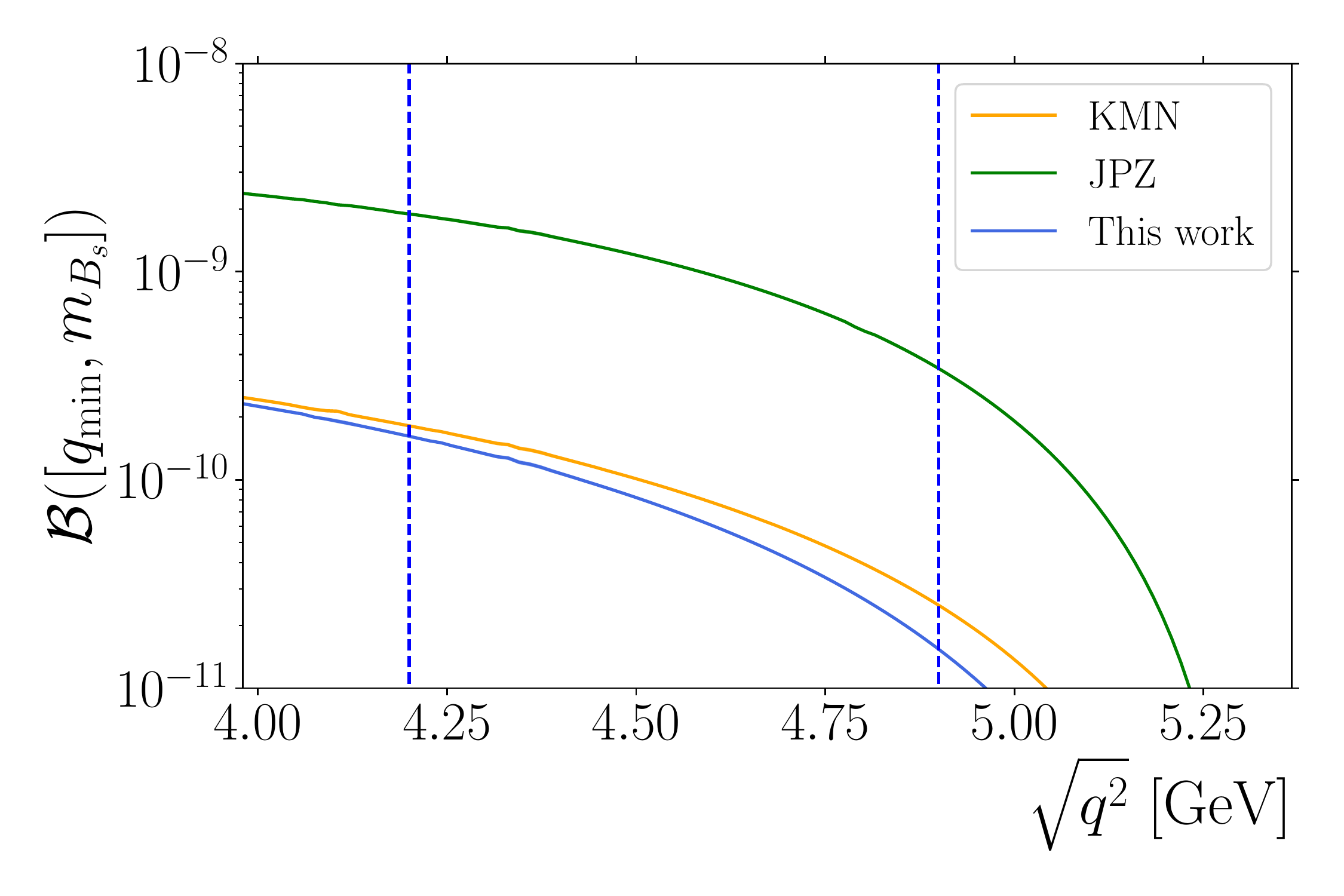}
\centering
\caption{Differential (top) and integrated (bottom) branching fraction of \Bsmmy. Color codes and vertical lines as in fig.~\ref{fig:KMN-vs-JPZ-vs-GNSV}. Left and right panels include only FF and respectively charmonium-resonance uncertainties.}
\label{fig:ff-vs-cc}
\end{figure}

A further remark concerns the question of the dominant component of the theory error. A breakdown of the uncertainties between FFs and charmonium resonances is provided in fig.~\ref{fig:ff-vs-cc}, for the differential and integrated branching fractions. The FF uncertainties are seen to be largely dominant over the uncertainties induced by the modeling of charmonium resonances, as was already observed in previous work \cite{Guadagnoli:2017quo,Carvunis:2021jga}. Since charmonium resonances escape any rigorous treatment, it is fortunate that, in this region, their contribution is small. In turn, while FF uncertainties are still large, a first-principle approach to their calculation exists, at least in this region, hence their error is reducible.
Finally, in table \ref{tab:int-BR} we provide the predictions and uncertainties of the branching fraction in the range $[4.2~\GeV,m_{\Bs}]$ for the three parametrizations considered.

\begin{table}[h!]
\renewcommand{\arraystretch}{1.5}
\begin{center}
\begin{tabular}{|c|c|}
  \hline
  \multicolumn{2}{|c|}{\gcl $\mc B(\bsmumugamma)[4.2~\GeV, m_{\Bs}]$} \\
  \hline
this work & $( 1.63  \pm  0.80) \times 10^{-10}$ \\ 
KMN \cite{Kozachuk:2017mdk} & $( 1.83 \pm 0.69  ) \times10^{-10}$ \\ 
JPZ \cite{Janowski:2021yvz} & $(1.90 \pm 0.53) \times10^{-9}$ \\ %
\hline
\multicolumn{2}{|c|}{\gcl \begin{Tabular}[1]{c} Influence of the choice of \FTVA \\ (with \FVA from this work) \end{Tabular}} \\
  \hline 
  $\FTVA$ from KMN & $(1.22 \pm 0.70) \times 10^{-10}$ \\ 
  $\FTVA$ from JPZ & $(0.92 \pm 0.58) \times 10^{-10}$ \\ 
  $\FTVA = 0$      & $(1.63 \pm 0.80) \times 10^{-10}$ \\ 
  \hline
\end{tabular}
\end{center}
\renewcommand{\arraystretch}{1.0}
\caption{(Top) Integrated branching fraction in the $[4.2~\GeV,m_{\Bs}]$ range for the three $B_s \to \gamma$ FF parametrizations discussed in the text. (Bottom) Influence on the integrated branching ratio of the choice of the tensor FFs between either the KMN or the JPZ parametrizations, or neglecting these FFs ($\FTVA = 0$).}
\label{tab:int-BR}
\end{table}

\begin{figure}[h!]
\centering
\includegraphics[width=0.49\textwidth]{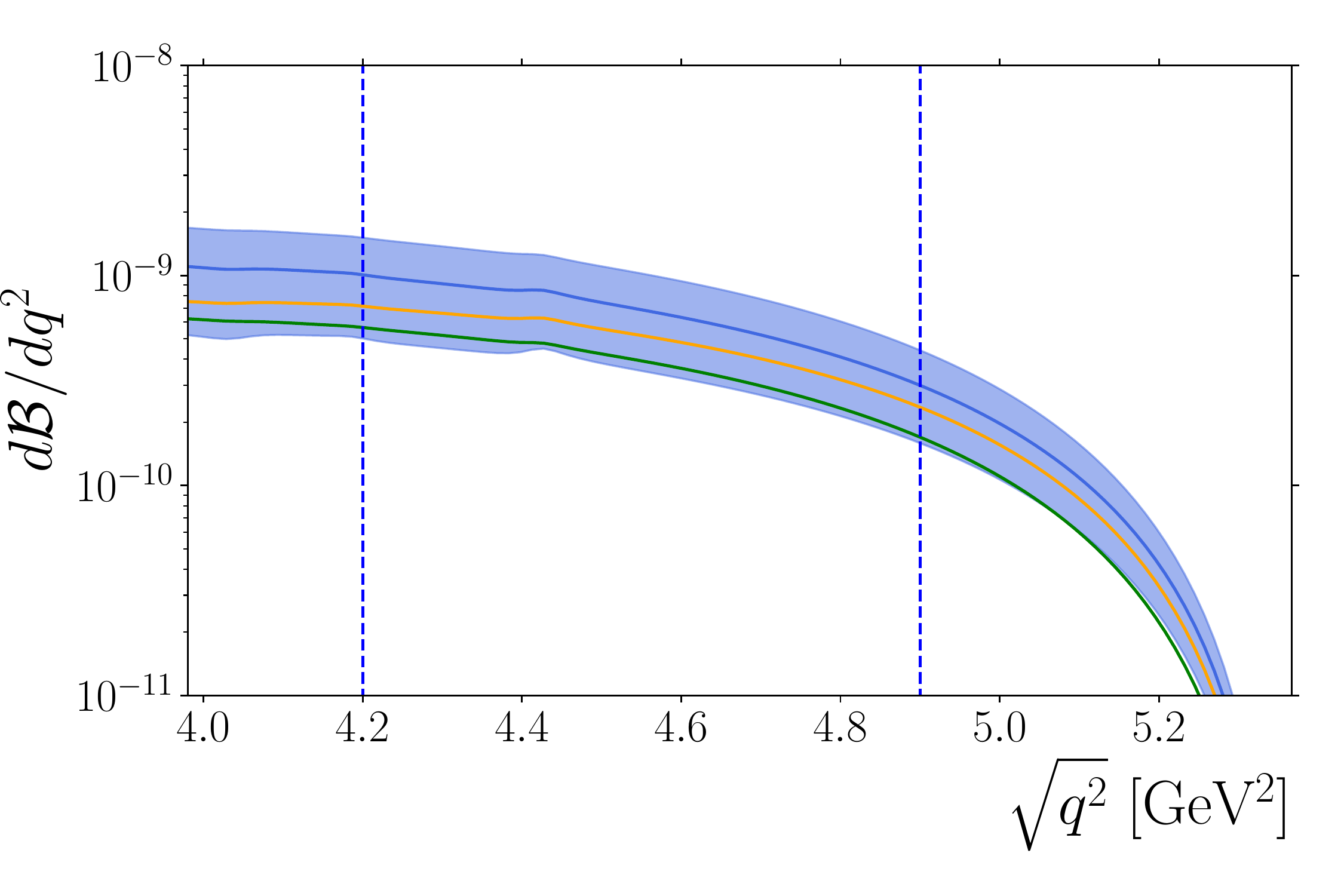}
\includegraphics[width=0.49\textwidth]{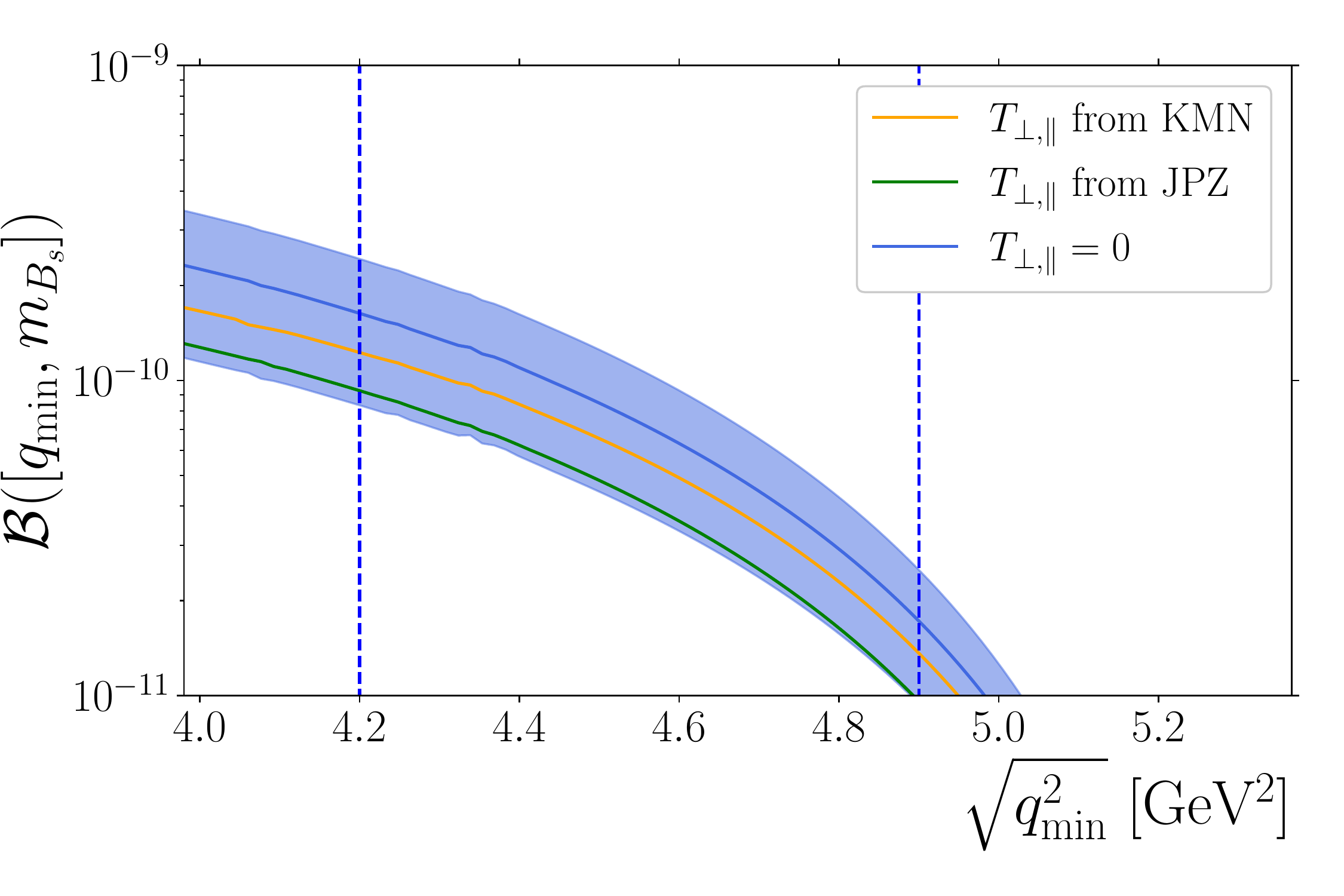}
\centering
\caption{Integrated $B_s \to \mu^+ \mu^- \gamma$ branching fraction as a function of $q^2_{\rm min}$ (see eq.~(\ref{eq:int-BR})), using \FVA from this work, and \FTVA from either KMN \cite{{Kozachuk:2017mdk}}, or JPZ \cite{Janowski:2021yvz}, or set to zero, see also legend.}
\label{fig:Int-BR-FTVA}
\end{figure}
  
\section{Conclusions} \label{sec:conclusions}

We provide a new estimate of the vector and axial form factors for $B_s \to \gamma$, which constitute the most important theoretical input for the prediction of $\mc B(B_s \to \mu^+ \mu^- \gamma)$ at high $q^2$, whose measurement via the indirect method is anticipated, following the recent LHCb limit \cite{LHCb:2021vsc,LHCb:2021awg}.

For this estimate, we adopt an approach that uses $D_s \to \gamma$ form factors directly computed on the lattice and scales them up to their $B_s$ counterparts using a suitable parameterization whose dependence on the heavy-quark mass is well-established. Our approach has three main premises: vector-meson dominance, which is expected to hold in the high-$q^2$ region of interest to us, in the form of an expansion of the spectral density into one {\em or more} physical poles; the relation between the residues of the poles and the effective coupling, or tri-coupling, between the appropriate vector meson, the ground-state pseudoscalar, and the photon; the parameterization of the tri-coupling in terms of quark magnetic moments.

Although each of these hypotheses is phenomenological, as opposed to first-principle, we seek validation through a number of cross-checks discussed in the text, which return a quite coherent picture. One reason why our approach has chances of being reliable is that the extrapolation is in the direction charm $\to$ bottom. In other words, we expect that our procedure would have been much less reliable in the opposite direction, namely if the lattice-QCD data were in the bottom sector and we had to extrapolate them to the charm sector.

We use our inferred $B_s \to \gamma$ form factors to reappraise the theory prediction for $\mc B(B_s \to \mu^+ \mu^- \gamma)$ in the range $\sqrt{q^2} \in [4.2~\GeV, m_{\Bs}]$, or subranges thereof, which represents a likely window for the experimental measurement. Our results are summarized in fig.~\ref{fig:KMN-vs-JPZ-vs-GNSV} and table~\ref{tab:int-BR}.

The validation of our approach rests ultimately in a first-principle calculation of the $B_s \to \gamma$ form factors, e.g. the $B_s$ counterpart of the calculation in Ref.~\cite{Desiderio:2020oej}. This being spelled out, our approach lends itself to certain well-defined lines of development.

First and foremost, our approach may be made more systematic, by careful inclusion, in the description of the relevant hadronic form factors, of basic properties such as analiticity, unitarity and the general form expected for the dispersion relation \cite{Okubo:1971jf, Okubo:1971my, Bourrely:1980gp, Lellouch:1995yv, Caprini:1997mu, Boyd:1997kz}. Possible avenues in this respect include the recent Dispersion-Matrix method of Refs.\,\cite{DiCarlo:2021dzg, Martinelli:2021frl}, which has been recently applied to many charged-current semileptonic transitions~\cite{Martinelli:2021onb, Martinelli:2021myh, Martinelli:2022tte, Martinelli:2022xir}, or the dispersive-bounds approach of Refs.\,\cite{Bobeth:2017vxj, Gubernari:2020eft}, deployed for several neutral-current semi-leptonic transitions~\cite{Gubernari:2022hxn, Blake:2022vfl,Amhis:2022vcd}. The question is whether a similar theoretical framework may be also applicable to hadronic FFs entering in rare-and-radiative meson decays such as those considered in the present work. This question is particularly relevant in view of lattice determinations of the FFs entering in $B_s \to \gamma$ decays, since it is not guaranteed that these data will cover the whole physical kinematical region $\sqrt{q^2} \in [2 m_\mu, m_{\Bs}]$. 

\section*{Acknowledgments}

We acknowledge useful discussions and correspondence with Damir Be\v{c}irevi\'{c} and Roman Zwicky. This work is supported by ANR under contract n. 202650 (GammaRare).

\appendix 

\section{Meson decay constants}
\label{app:meson_decay_constants}

\begin{figure}[b]
\centering
\includegraphics[width=0.49\textwidth]{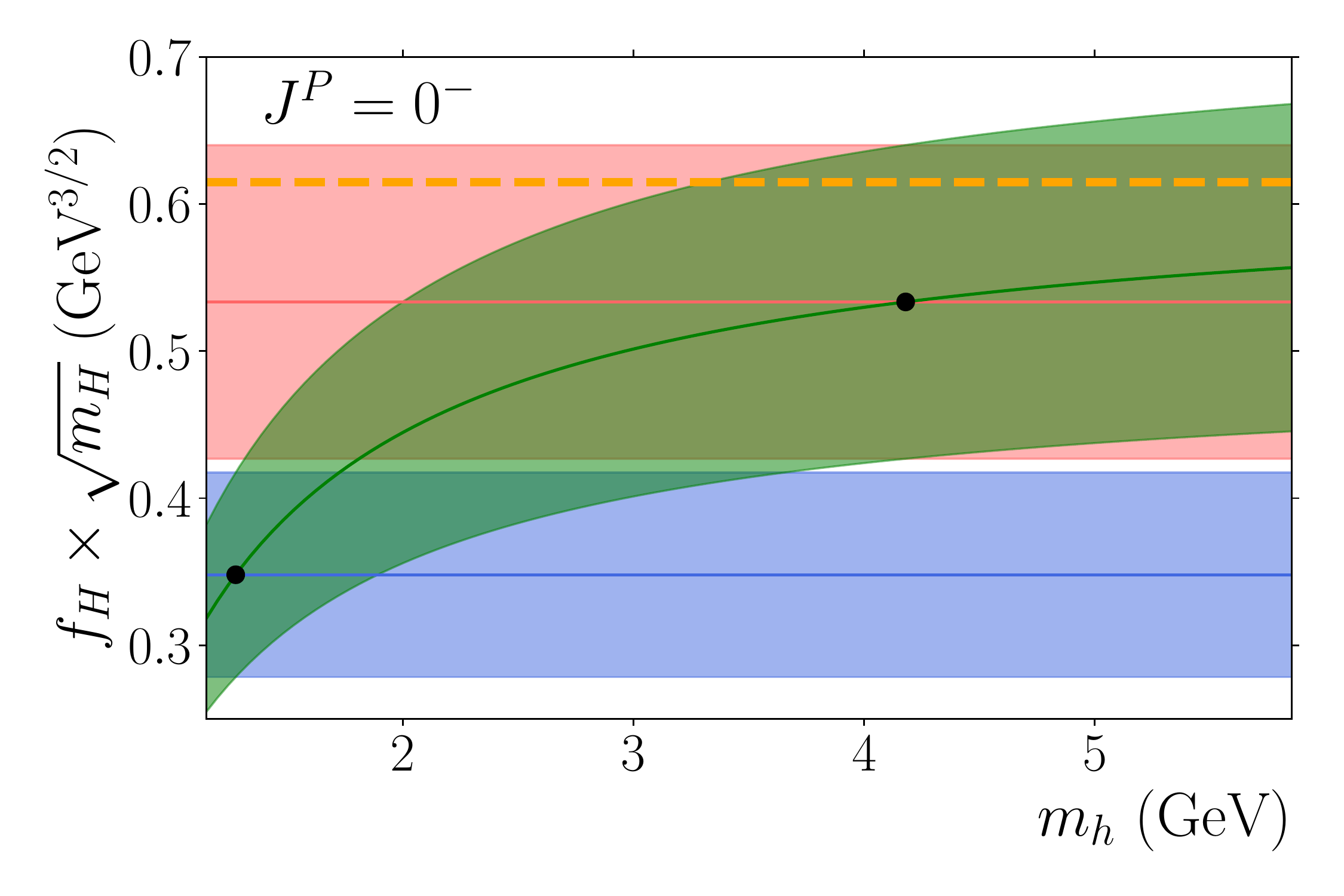}
\includegraphics[width=0.49\textwidth]{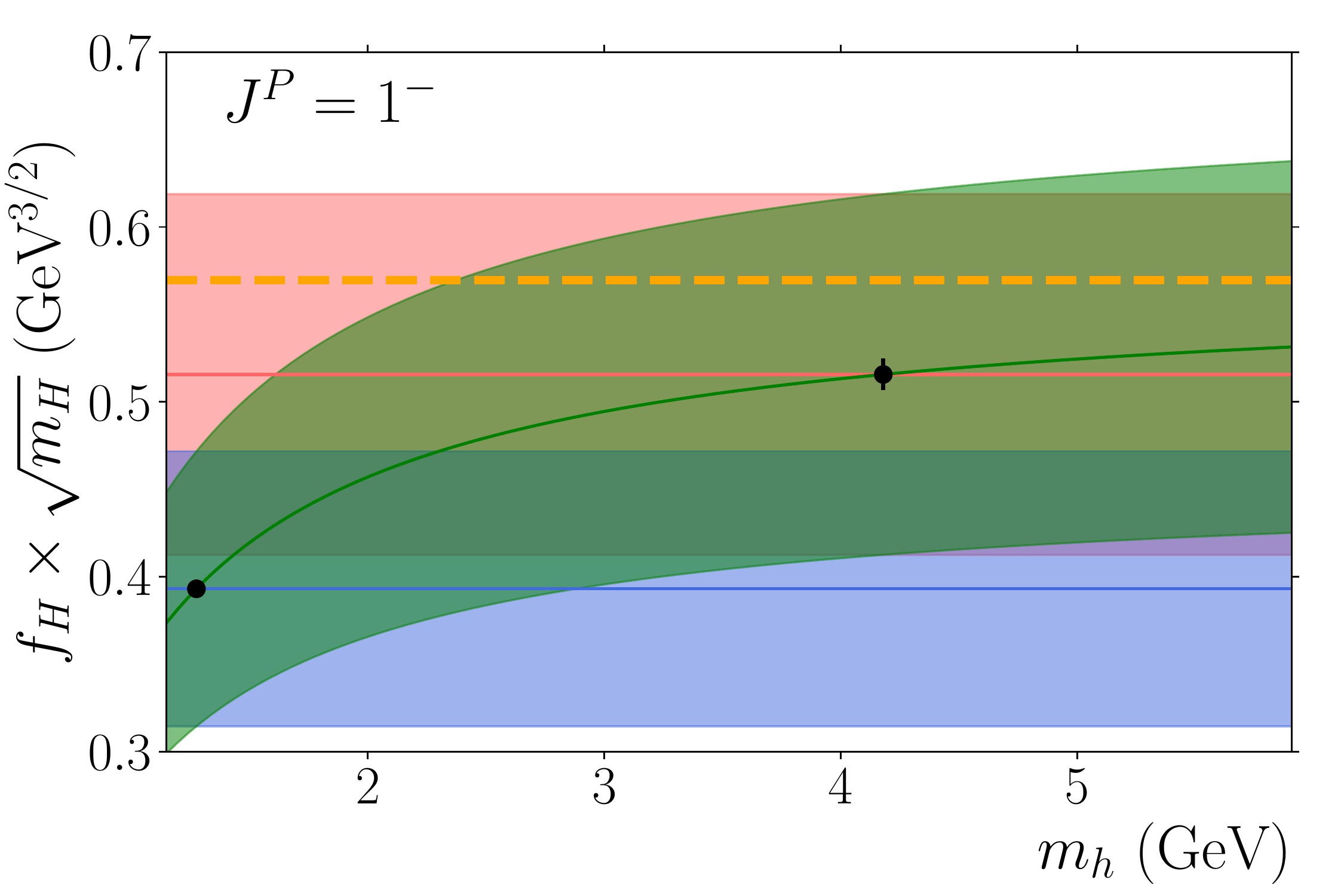}
\includegraphics[width=0.49\textwidth]{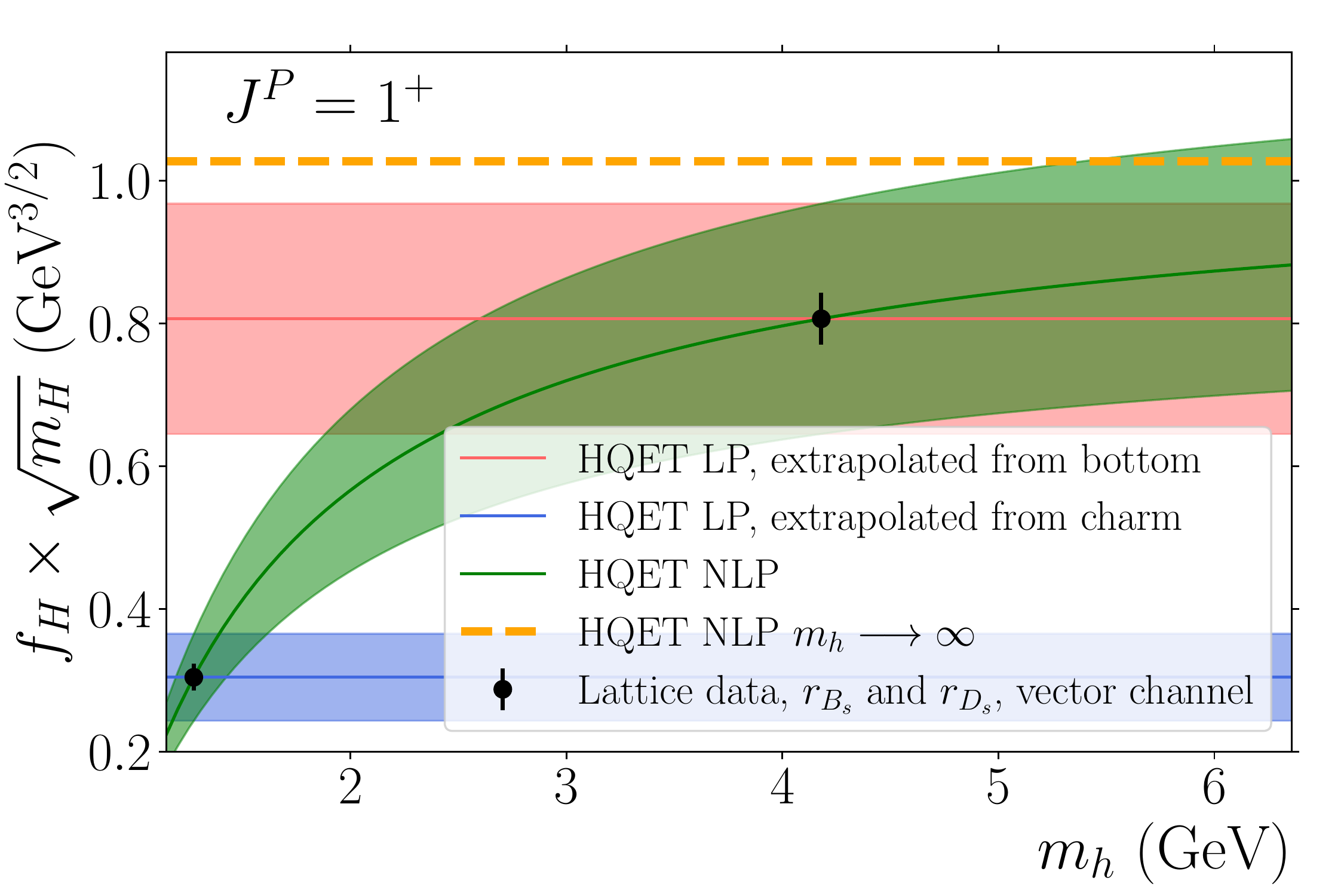}
\centering
\caption{
Product $\sqrt{m_H}\times f_H$ as a function of the heavy quark mass, (top left) for the pseudoscalar $0^-$ meson, (top right) vector meson $1^-$, and (bottom) axial-vector meson $1^+$. The red bands correspond to the predictions of leading-power (LP) HQET using the bottom sector computations of LQCD, the blue bands using the charm sector computations, and the green bands represent the NLP HQET expansion using computations from both sectors.}
\label{fig:scaling-law}
\end{figure}

The values of, and references for, the meson decay constants used in this work are summed up in table \ref{tab:decay-constants}, along with the computational methods employed. For the $D_{s1}^*$ and $B_{s1}^*$ states we are only aware of quark-model computations. One could argue that it would be possible to access such decay constants using a scaling law similar to the one applied to the tri-couplings, and discussed in Sec. \ref{sec:g=QxMM}. We discuss hereafter why such a procedure would not lead to an accuracy satisfactory for the present study.
\begin{table}[h]
\renewcommand{\arraystretch}{1.5}
\begin{center}
\begin{tabular}{|c|cccc|}
\hline
\gcl $J^P$ & \multicolumn{4}{c|}{\gcl Charm sector, in MeV} \\
\hline
\hline
$0^-$ & $f_{D_s}$      & $249.9(0.5)$ & \cite{Bazavov:2017lyh,Carrasco:2014poa,FlavourLatticeAveragingGroupFLAG:2021npn}  & LQCD $N_f = 2+1+1$ \\
$1^-$ & $f_{D_s^*}$     & $270.5(2.9)$   &  \cite{Donald:2013sra,Lubicz:2017asp,Blossier:2018jol, Balasubramamian:2019wgx,Chen:2020qma} & LQCD \\
$1^-$ & $f_{D_{s1}^*}$   & $101(20)$ & \cite{Devlani:2011zz, Devlani:2012zz} & Quark Model                                                      \\
$1^+$ & $f_{D_{s1}}$   & $194(12)$  &  \cite{Bali:2017pdv}  &  LQCD \\
\hline
\hline
\gcl $J^P$ & \multicolumn{4}{c|}{\gcl Bottom sector, in MeV } \\
\hline
\hline
$0^-$ & $f_{B_s}$      & $230.3(1.3)$ & \cite{Bazavov:2017lyh,ETM:2016nbo,Dowdall:2013tga,Hughes:2017spc,FlavourLatticeAveragingGroupFLAG:2021npn} & LQCD $N_f = 2+1+1$  \\
$1^-$ & $f_{B_s^*}$     & $221.6(3.9)$ &  \cite{Colquhoun:2015oha,Lubicz:2017asp,Balasubramamian:2019wgx}   & LQCD \\
$1^-$ & $f_{B_{s1}^*}$   & $107(21)$  &  \cite{Devlani:2011zz, Devlani:2012zz} & Quark Model \\
$1^+$ & $f_{B_{s1}}$   & $334(15)$  &  \cite{Wang:2015mxa,Pullin:2021ebn} & LCSR  \\
\hline
\end{tabular}
\end{center}
\renewcommand{\arraystretch}{1.0}
\caption{Inputs for the meson decay constants, in MeV. The quoted value is the weighted average of all determinations in the references provided. The values of $f_{D_s}$ and $f_{B_s}$ are exactly the ones reported in Ref.~\cite{FlavourLatticeAveragingGroupFLAG:2021npn}. The quoted value for $f_{B_s^*}$ follows from the computation of the ratio $f_{B_s^*}/f_{B_s}$in Ref.~\cite{Balasubramamian:2019wgx} plus the value of $f_{B_s}$ averaged in~\cite{FlavourLatticeAveragingGroupFLAG:2021npn}. An uncertainty of 20\% is associated to the Quark-Model computation of Refs.~\cite{Devlani:2011zz, Devlani:2012zz}. As discussed in the text, this figure is used {\em for reference only}, i.e. to test whether this determination agrees with the scaling law within 20\% or not. The rightmost column describes the method used.}
\label{tab:decay-constants}
\end{table}
The scaling of the meson decay constants is often referred to be $\sim m_H^{-1/2}$, with $m_H$ the mass of a heavy meson $H$. This scaling originates from the normalization of a heavy meson $H$ with velocity $v$ in heavy-quark effective theory (HQET \cite{Georgi:1990um}, for reviews see \cite{Neubert:1993mb,Manohar:2000dt}),
\be
\label{eq:Goldstone-theorem}
\bra{0}\bar{q}\gamma^\mu\gamma^5 Q\ket{H}_\text{HQET} = ia v^\mu = \frac{1}{\sqrt{2m_H}}\bra{0}\bar{q}\gamma^\mu\gamma^5 Q\ket{H}_\text{QCD} ~,
\ee
where the r.h.s. refers to the ordinary relativistic matrix element. Neglecting short-distance corrections \cite{Neubert:1992fk}, the constant $a$ does not depend on the heavy flavor, and thus yields the scaling law\footnote{The decay constant is defined as $\bra{0}\bar{q}\gamma^\mu\gamma^5 Q\ket{H} \equiv if_Hp^\mu$ following  Ref.~\cite{FlavourLatticeAveragingGroupFLAG:2021npn}. The overall sign on the r.h.s. is relevant for the correct sign of the interference term in $\mc B(\Bsmmy)$, see beginning of Sec.~\ref{sec:preliminaries}.}
\be
\label{eq:a-def}
f_H \propto \frac{a}{\sqrt{m_H}} ~.
\ee
One could then hope to use such a scaling law to relate the decay constants of excited meson states differing by the heavy-quark flavour, e.g. the $\Ds$ and the $\Bs$. The accuracy one may expect from such a strategy can be put to the test using the latest LQCD determinations of the decay constants. Referring to table~\ref{tab:decay-constants} for the numerical values, one can namely compare the values of the product $\sqrt{m_H}\times f_H$ for different heavy flavored mesons. The results are plotted in fig.~\ref{fig:scaling-law} for $H=\Ds,\Bs$, in the pseudoscalar, vector and axial-vector cases, assigning a 20\% uncertainty to the scaling law {\em for reference only}. We see that the values predicted by the scaling law do not agree within this reference 20\% level for the pseudoscalar and axial-vector cases suggesting that, in order to reproduce LQCD data, one would need further power-suppressed terms in the HQET expansion. We can estimate the relative size $b / m_h$ of such terms in the case of the pseudoscalar and the first pole of the vector and axial channels by fitting the respective LQCD data to
\be \label{eq:NLP-exp}
f_H\times\sqrt{m_H} = a\left(1-\frac{b}{m_h}\right) ~,
\ee
where, again, we neglect short-distance corrections \cite{Neubert:1992fk}.
Note that $m_h$ denotes the heavy-quark mass $m_h$ (a separate entity than $m_H$) and that we use the kinetic scheme for the quark masses. The results are summarized in table~\ref{tab:decay-constant-expansion}. The LP contributions to the pseudoscalar and vector channels agree with each other, but not with the axial channel. The latter appears to require a subleading-power correction to the decay-constant scaling law as large as 70\% at the charm mass, and 20\% at the bottom mass, whereas the corresponding corrections in the pseudoscalar and vector channel are somewhat more contained. The consistency between the $a$ values in the pseudoscalar and vector channels, and their difference with the axial-channel value are entirely expected, and so are the sizes of the power-suppressed corrections. For instance, Ref.~\cite{Lubicz:2017asp} studied the analogous scaling of the ratio between the vector and the pseudoscalar, finding power-suppressed corrections of about 30\% at the charm mass and 3\% at the bottom mass. Such corrections (to ratios) are then directly comparable with the {\em difference} between the corresponding corrections in table \ref{tab:decay-constant-expansion}. This comparison shows very good consistency at the bottom mass. At the charm mass we observe a more contained correction, of $(43-30)\% = 13\%$, but we should keep in mind that our eq.~(\ref{eq:NLP-exp}) does not include terms of order $(1 / m_h)^2$, that are instead accounted for in the parameterization of Ref.~\cite{Lubicz:2017asp}. Such corrections will change our $b/m_c$ values by at least $O(10\%)$.

\begin{table}[h!]
\renewcommand{\arraystretch}{1.5}
\begin{center}
\begin{tabular}{|c|c|c|c|}
\hline
\gcl Parameters & \gcl Pseudoscalar  &  \gcl Vector  &  \gcl Axial-vector  \\
\hline
$a$~$(\SI{}{GeV^{3/2}})$                                      & $0.61$ & $0.56$ & $1.03$ \\
$b$ (MeV)        & $552$ & $394$  & $896$ \\
$\frac{b}{m_c}$  & $43\%$ & $ 30\%$  &  $70\%$ \\
$\frac{b}{m_b}$  & $13\%$ & $ 10\%$   & $21\%$  \\
\hline
\end{tabular}
\end{center}
\renewcommand{\arraystretch}{1.0}
\caption{Parameters of an HQET-inspired expansion of the product $f_H\times\sqrt{m_H}$ according to eq.\,(\ref{eq:NLP-exp}). The last two lines quantify the importance of next-to-leading-power corrections in the $c$ and $b$ case, respectively.}
\label{tab:decay-constant-expansion}
\end{table}

\section{Table of inputs}\label{app:input_table}

In table \ref{tab:input} we collect all our inputs. None of the parameters listed contributes a non-negligible part of the theory uncertainties in $\mc B(\Bsmmy)$, with the exception of the broad-charmonium phases $\delta_V$, whose treatment is discussed in Sec. \ref{sec:broad_cc}. Any omitted parameter is taken from Ref. \cite{Beneke:2020fot}. The CKM input is taken from the latest `global Standard-Model analysis' and `New-Physics fit', which are available from ~\cite{UTfit-website,Ciuchini:2000de,UTfit:2022hsi,Bona:2022zhn}. Similar results may be obtained from \cite{CKMfitter-website,Charles:2004jd}.
\noindent \begin{table*}[h]
\def\arraystretch{1.4}
\begin{center}
\begin{tabular}{|c|c|c||c|c|c|}
\hline
\gcl Parameter & \gcl Value & \gcl Ref. & \gcl Parameter & \gcl Value & \gcl Ref. \\ 
\hline
\hline
$m_{\Bs}$ & $ 5.36688~\GeV$  & \multirow{2}{*}{\cite{ParticleDataGroup:2022pth}} & $\alpha_s (m_Z)$ & $0.1179$ & \multirow{3}{*}{\cite{ParticleDataGroup:2022pth}} \\
$\tau_{Bs}$ & $ 1.520\times10^{-12}~$s  & & $\Delta\Gamma_{\Bs}$ & $0.084\times10^{-12}~$s  & \\ 
$f_{Bs}$ & $ 0.2303~\GeV$  & \cite{FlavourLatticeAveragingGroup:2019iem} & $\alpha_{\rm e.m.}(m_b)$ & $1/132.1$ & \\ \hline
$m_b(m_b)$ & $ 4.18~\GeV$ & \multirow{4}{*}{\cite{ParticleDataGroup:2022pth}} & $\lambda$ & $ 0.225$ & \multirow{4}{*}{\cite{UTfit-website,UTfit:2022hsi}} \\
$m_c(m_c)$ & $ 1.27~\GeV$ & & $A$ & $0.828$ & \\
$m_{b}^{\rm pole}$ & $ 4.78~\GeV$ & & $\bar{\rho}$ & $0.160$ & \\
$m_c^{\rm pole}$ & $ 1.67~\GeV$ & & $\bar{\eta}$ & $0.347$ & \\ \hline
$m_{\psi(2S)} $ & $3.686~\GeV$ & \multirow{5}{*}{\cite{ParticleDataGroup:2022pth}} & $\Gamma_{\psi(2S)}$ & $ 0.294 \times 10^{-3}~\GeV$ & \multirow{5}{*}{\cite{ParticleDataGroup:2022pth}} \\
$m_{\psi(3770)} $ & $3.774~\GeV$ & & $\Gamma_{\psi(3770)}$ & $ 27.2 \times 10^{-3}~\GeV$ & \\
$m_{\psi(4040)} $ & $4.039~\GeV$ & & $\Gamma_{\psi(4040)}$ & $ 80 \times 10^{-3}~\GeV$ & \\
$m_{\psi(4160)} $ & $4.191~\GeV$ & & $\Gamma_{\psi(4160)}$ & $ 70 \times 10^{-3}~\GeV$ & \\
$m_{\psi(4415)} $ & $4.421~\GeV$ & & $\Gamma_{\psi(4415)}$ & $ 62 \times 10^{-3}~\GeV$ & \\ \hline
$\mathcal{B}(\psi(2S) \to \ell \ell)$ & $ 8.0\times 10^{-3} $ & \multirow{5}{*}{\cite{ParticleDataGroup:2022pth}} & $\delta_{\psi(2S)}$ & $0 $  & \multirow{5}{*}{\cite{BES:2007zwq}} \\
$\mathcal{B}(\psi(3770) \to \ell \ell)$ & $ 9.6 \times 10^{-6} $ & & $\delta_{\psi(3770)}$ & $0 $  & \\
$\mathcal{B}(\psi(4040) \to \ell \ell)$ & $ 10.7 \times 10^{-6} $ & & $\delta_{\psi(4040)}$ & $133 \times \pi/180 $  & \\
$\mathcal{B}(\psi(4160) \to \ell \ell)$ & $ 6.9 \times 10^{-6} $ & & $\delta_{\psi(4160)}$ & $301 \times \pi/180 $  & \\
$\mathcal{B}(\psi(4415) \to \ell \ell)$ & $ 2.0 \times 10^{-5} $ & & $\delta_{\psi(4415)}$ & $246 \times \pi/180 $  & \\
\hline
\end{tabular}
\caption{List of input parameters.}
\label{tab:input}
\end{center}
\def\arraystretch{1.4}
\end{table*}

\bibliography{bibliography}
\bibliographystyle{JHEP}

\end{document}